\documentstyle[eqsecnum,prd,aps,twocolumn,epsf,floats,amsfonts,amssymb,amsmath]{revtex}
\newcommand{\bea}{\begin{eqnarray}}
\newcommand{\eea}{\end{eqnarray}}
\newcommand{\ba}{\begin{array}}
\newcommand{\ea}{\end{array}}

\newcommand{\bit}{\begin{itemize}}
\newcommand{\eit}{\end{itemize}}
\newcommand{\ben}{\begin{enumerate}}
\newcommand{\een}{\end{enumerate}}

\newcommand{\noi}{\noindent}

\begin{document}

\twocolumn[\hsize\textwidth\columnwidth\hsize\csname@twocolumnfalse\endcsname

\typeout{--- Title page start ---}

\begin{flushright}
{\tt cond-mat/0207707}\\
\end{flushright}
\vskip 3mm

\title{The world according to R{\'e}nyi:
Thermodynamics of multifractal systems}
\author{Petr Jizba${}^{}$ and
Toshihico Arimitsu${}^{}$}

\address{ $ $\\[2mm]
${}^{}$ Institute of Physics, University of Tsukuba, Ibaraki
305-8571, Japan
\\ [2mm] E-mails: petr@cm.ph.tsukuba.ac.jp, arimitsu@cm.ph.tsukuba.ac.jp
 }

\vspace{3mm}


\maketitle

\begin{center}
{\small \bf Abstract}
\end{center}
\begin{abstract}
\noindent We discuss basic statistical properties of systems with
multifractal structure. This is possible by extending the notion
of the usual Gibbs--Shannon entropy into more general framework -
R{\'e}nyi's information entropy. We address the renormalization
issue for R{\'e}nyi's entropy on (multi)fractal sets and
consequently show how R{\'e}nyi's parameter is connected with
multifractal singularity spectrum. The maximal entropy approach
then provides a passage between R{\'e}nyi's information entropy
and thermodynamics of multifractals. Important issues such as
R\'{e}nyi's entropy versus Tsallis--Havrda--Charvat entropy and
PDF reconstruction theorem are also studied. Finally, some further
speculations on a possible relevance of our approach to cosmology
are discussed.
\\

\vspace{3mm}
\end{abstract}
\vskip5mm]



\section{Introduction}
 The past two decades have witnessed an explosion of activity and
progress in both equilibrium and non--equilibrium statistical
physics. The catalyst has been the massive infusion of ideas from
information theory, theory of  chaotic dynamical systems, theory
of critical phenomena, and quantum field theory. These ideas
include the generalized information measures, quasi--periodic and
strange attractors, fully developed turbulence, percolation,
renormalization of large--scale dynamics,  and attractive, albeit
speculative, ideas about  quark--gluon plasma formation and
dynamics. It is the purpose of this paper to proceed in this line
of development. The issue at the stake is what modifications in
statistical physics one should expect when dealing with systems
with varied fractal dimension - multifractals. The view which we
present here hinges on two mutually interrelated concepts, namely
on R{\'e}nyi's information entropy\cite{Re1,Re2} and
(multi)fractal geometry. In this connection we would like to
stress that in order to exhibit the link between R\'{e}nyi
information entropies and (multi)fractal systems as generally as
possible we do not put much emphasize on the important yet rather
narrow class of (multi)fractal systems - chaotic dynamical
systems.

\vspace{3mm}

One of the fundamental observations of information theory is that
the most general functional form for the mean transmitted
information (i.e., information entropy) is that of R{\'e}nyi. In
Section II we briefly outline R{\'e}nyi's proof and discuss some
fundamentals from information theory which will show up to be
indispensable in following sections. We show that with certain
mathematical cautiousness Shannon's entropy can be viewed as a
special example of R{\'e}ny's entropy in case when R{\'e}nyi's
parameter $\alpha \rightarrow 1$. We also address the question of
the status of Tsallis--Havrda--Charvat (THC)
entropy\cite{HaCh,Ts1} in the framework of information theory.

\vspace{3mm}

Although R{\'e}nyi's information measure offers very natural - and
maybe conceptually the cleanest - setting for the entropy, it has
not found so far as much applicability as Shannon's (or Gibbs's)
entropy. The explanation, no doubt, lies in two facts; ambiguous
renormalization of R{\'e}nyi's entropy for non--discrete
distributions and little insight into the meaning of R{\'e}nyi's
$\alpha$ parameter. Surprisingly little work has been done towards
understanding both of the former points. In Section III we aim to
address the first one. We choose, in a sense, a minimal
renormalization prescription conforming to the condition of
additivity of independent information. R{\'e}nyi's entropy thus
obtained is then directly related to the information content
(``negentropy").

\vspace{3mm}

To clarify the position of R{\'e}nyi's entropy in physics, or in
other word, to find the physical interpretation for $\alpha$
parameter, we resort in Section IV to systems with a multifractal
structure. Such systems are very important and highly diverse,
including  the turbulent flow of fluids\cite{Pa1,ArAr1},
percolations\cite{Ah1}, diffusion--limited aggregation (DLA)
systems\cite{NST1}, DNA sequences\cite{ZGU1}, finance\cite{VVA1},
and string theory\cite{MMW1}. Using the reconstruction theorem we
argue that in order to obtain a ``full" information about a
(multi)fractal system  we need to know R{\'e}nyi's entropies to
all orders. Still, for discrete spaces and simple metric spaces
(like ${\mathbb{R}}^d$) we find that the contribution from
Shannon's entropy dominates over all other R{\'e}nyi entropies. We
further show that from the maximal entropy (MaxEnt) point of view,
extremizing the Shannon entropy on a multifractal is equivalent to
extremizing directly Renyi's entropy without invoking the
multifractal structure explicitly. Application of this result to a
cosmic strings network will be presented elsewhere\cite{AJS1}.


\vspace{3mm}



We close with Section V where we present some speculations on the
relevance of the outlined approach to string cosmology and quantum
mechanics.
%
For reader's convenience we supplement the paper with eight
appendices which clarify some finer mathematical manipulations.

\section{Renyi's entropy of discrete probability distributions}

\subsection{R{\'e}nyi's entropy and information theory}

We begin this section by summarizing the information theory
procedure leading to R{\'e}nyi's entropy\cite{Re1,Re2}. This is of
course well known but it may be useful to repeat it here in order
to make our discussion self--contained. We will also need to
generalize it when considering THC entropy in Section IID and
axiomatization of R\'{e}nyi's entropy in Appendix B .

\vspace{3mm}

Let us start with a discrete probability distribution ${\cal{P}} =
\{ p_1, p_2, \ldots, p_n \}$ fulfilling usual conditions
\begin{equation}
p_k \ge 0\, , \;\;\; \sum_k p_k =1 \, .
\end{equation}
\noindent We then assume three things about information. Firstly,
information should be additive for two independent events.
Secondly, information should purely depend on ${\cal{P}}$.  These
two condition can be also formulated in the following way: if we
observe the  outcome of two independent events with respective
probabilities $p$ and $q$, then the total received information is
the sum of two partial ones. Therefore the following functional
equality holds:
\begin{equation}
{\cal{I}}(pq) = {\cal{I}}(p)  + {\cal{I}}(q)\, .
\end{equation}
\noindent The latter is well known modified Cauchy's functional
equation\cite{Acz1} which has (under fairly broad
assumptions\cite{Re2,Rb1}) unique class of solutions -
$\kappa\,\mbox{log}_2(\ldots)$. The constant $\kappa$  is then
fixed via appropriate ``boundary'' condition.
Setting
${\cal{I}}(1/2) = 1$ we obtain the, so
called, {\em Hartley measure} of information\cite{Ha1}. So the
amount of information received by learning that event of
probability $p$ took place equals
\begin{equation}
{\cal{I}}(p) = - \mbox{log}_2 (p)\, .
\end{equation}
\noindent The third assumption is that if different amounts of
information occur with different  probabilities, the total amount
of information is the average of the individual  information
weighted by the probabilities of their occurrences. In general, if
the possible outcomes of an experiment
are
${\cal{A}}_1, {\cal{A}}_2, \ldots, {\cal{A}}_n$ with corresponding
probabilities $p_1, p_2, \ldots, p_n$, and ${\cal{A}}_k$ conveys
${\cal{I}}_k$ bits of information, then the total amount of
information conveyed would be
\begin{equation}
{\cal{I}}({\cal{P}},\Im ) = \sum_{k=1}^{n}p_k {\cal{I}}_k \, ,
\label{I5}
\end{equation}
\noindent where $\Im = \{ {\cal{I}}_1, {\cal{I}}_2, \ldots,{\cal{I}}_n \}$.
However, the linear averaging implemented in (\ref{I5}) is only a
specific case of a more general
mean. If $f$ is a real function having an inverse $f^{-1}$ then the
number
\begin{equation}
f^{-1}\left(\sum_{k}^n  p_k f(x_k) \right) \, ,
\label{I6}
\end{equation}
\noindent is called the mean value of $x_1, x_2, \ldots, x_n$
associated with $f$. As shown in Refs.\cite{Ko1,Na1,Ac1},
(\ref{I6}) prescribes the most general mean compatible with
postulates of probability theory (see, eg., \cite{Re1}). The
function $f$ is often referred to as Kolmogorov--Nagumo's function
.


\vspace{3mm}

Former analysis suggests that in the most general case the measure
of the amount of transmitted information should admit the form
\begin{equation}
{\cal{I}}({\cal{P}},\Im ) = f^{-1}\left(\sum_{k=1}^{n}p_k\, f
\left(-\mbox{log}_2(p_k)\right) \right) \, .
\label{I7}
\end{equation}
\noindent The natural question arises, what is the possible
mathematical form  of $f$, or in other words, what is the most
general class of functions $f$ which will still provide a measure
of information compatible with  the additivity postulate.
Obviously for a given set of outcomes, many possible means can be
defined, depending on which features of the outcomes are of
interest.
It comes therefore as a pleasant surprise to find that the
additivity postulate allows only for two classes of $f$'s - linear
and exponential functions. The proof of this statement is simple
and clarifies a good deal about  $f$ so for the future reference
we sketch its  main points. Alternative proof based on scaling
argumentation is presented in Appendix A.

\vspace{3mm}

Let an experiment ${\cal{K}}$ be a union of two independent
experiments ${\cal{K}}_1$ and   ${\cal{K}}_2$. Let further assume
that we receive ${\cal{I}}^{(1)}_k$ bits of information with
probability $p_k$ connected with ${\cal{K}}_1$ and
${\cal{I}}^{(2)}_l$  bits of information with probability $q_l$
connected with ${\cal{K}}_2$. As a result we receive
${\cal{I}}^{(1)}_k + {\cal{I}}^{(2)}_l$ bits of information with
probability $p_kq_l$. We assume further that there is $m$ possible
outcomes in ${\cal{K}}_1$ experiment (i.e., $k = 1,2, \ldots, m$)
and $n$ possible outcomes in ${\cal{K}}_2$ experiment (i.e., $l =
1,2, \ldots, n$). Invoking the postulate of additivity we may
write
\begin{eqnarray}
&&f^{-1}\left( \sum_{k}^{m}\sum_{l}^{n} p_k q_l \, f\left({\cal{I}}^{(1)}_k +
{\cal{I}}^{(2)}_l  \right) \right)\nonumber \\
&& \mbox{\hspace{0.15cm}} = f^{-1} \left( \sum_{k}^{m} p_k \,
f\left({\cal{I}}^{(1)}_k\right) \right) + f^{-1} \left( \sum_{l}^{n} q_l \,
f\left({\cal{I}}^{(2)}_l\right)\right)\, .
\label{I66}
\end{eqnarray}
\noindent The former must hold completely generally irrespective
of our choice of ${\cal{P}} = \{ p_1, \ldots, p_m\} $, ${\cal{Q}}
= \{ q_1, \ldots, q_n\} $ and irrespective of the actual choice of
independent experiments ${\cal{K}}_1$,  ${\cal{K}}_2$. So if we
choose ${\cal{I}}^{(2)}_l ={\cal{I}}$ independently of $k$ we
obtain from (\ref{I66})
\begin{eqnarray}
&&f^{-1}\left( \sum_{k}^{m} p_k \, f\left({\cal{I}}^{(1)}_k +
{\cal{I}}  \right) \right)\nonumber \\
&& \mbox{\hspace{2.5cm}}= f^{-1} \left( \sum_{k}^{m} p_k \,
f\left({\cal{I}}^{(1)}_k\right) \right) + {\cal{I}}\, .
\label{I67}
\end{eqnarray}
\noindent Let us denote $f_y (x) = f(x+y)$ (so namely $f^{-1}(x) -y =
f^{-1}_y (x)$). Thus Eq.(\ref{I67}) may be recast into the form
\begin{equation}
f^{-1}_{{\cal{I}}}\left( \sum_{k}^{m} p_k \, f_{{\cal{I}}}
\left({\cal{I}}^{(1)}_k \right) \right) = f^{-1}\left( \sum_{k}^{m} p_k \, f
\left({\cal{I}}^{(1)}_k \right) \right)\, .
\label{I677}
\end{equation}
\noindent So functions $f_{{\cal{I}}}$ and $f$ generate the same
mean. It is well known in theory of means (see eg.,\cite{Hardy1})
that Eq.(\ref{I677}) holds only if $f_{{\cal{I}}}$ is a linear
function of $f$. So we have
\begin{equation}
 f_{{\cal{I}}}(z) = f(z +{\cal{I}}) = a({\cal{I}})\,f(z) + b({\cal{I}})\, .
\label{fund1}
\end{equation}
\noindent Here $a(\ldots) \not= 0$ and $b(\ldots)$ are independent
of $z$. Without loss of generality we shall assume that $f(0) = 0$
(or otherwise we adjust $b$). As a result $b({\cal{I}}) =
f({\cal{I}})$. Therefore
\begin{eqnarray}
f(z + {\cal{I}}) &=&  a({\cal{I}})\,f(z) + f({\cal{I}})\nonumber \\
f(z + {\cal{I}}) &=&  a(z)\,f({\cal{I}}) + f(z)\, ,
\label{I666}
\end{eqnarray}
\noindent where the second line was obtained by a simple
interchange $z \leftrightarrow {\cal{I}}$. Subtraction of both
lines in (\ref{I666}) leads to the following separation of
variables ($z\not=0$, ${\cal{I}}\not=0$):
\begin{equation}
\frac{a(z) -1}{f(z)} =  \frac{a({\cal{I}}) -1}{f({\cal{I}})} = \gamma \, .
\label{I77}
\end{equation}
\noindent ($\gamma$ is a constant independent both of $z$ and ${\cal{I}}$).
The solution of (\ref{I77}) has a simple form
\begin{equation}
a(x) -1 = \gamma \, f(x)\, .
\label{I78}
\end{equation}
\noindent Note that (\ref{I78}) holds true also for $x=0$. In
connection with (\ref{I78}) it is useful to distinguish two cases;
$\gamma = 0$ and $\gamma \not= 0$. In the first case  when $\gamma
= 0$,  $a(x) =1$ and we get Cauchy's functional
equation\cite{Acz1}
\begin{equation}
f(z + {\cal{I}}) = f(z) + f({\cal{I}})\, ,
\end{equation}
\noindent which for $z, \mathcal{I} \in \mathbb{R}$ has the well
known solution: $f(x) = c\, x$ with the nonzero constant $c$. This
is in a sense the most elementary Kolmogorov--Nagumo function.
Plugging the latter into Eq.(\ref{I7}) the measure of transmitted
information boils down to Shannon's measure
\begin{equation}
{\cal{I}}({\cal{P}},\Im ) = -\sum_{k=1}^{n}p_k \,
\mbox{log}_2(p_k) \equiv \mathcal{H} \, .
\label{I79}
\end{equation}
\noindent In the second case when $\gamma \not= 0$, $a(x)$ fulfills
the modified Cauchy's functional equation\cite{Acz1}
\begin{equation}
a(z + {\cal{I}}) = a(z)a({\cal{I}})\, ,
\end{equation}
\noindent which for continuous $a(\ldots)$ and $z, \mathcal{I} \in
\mathbb{R}$ has only exponential solutions. Thus we may generally
write: $ a(x) = 2^{(1-\alpha)x}$ with $\alpha \not= 1$ being some
constants.  As a result we get $f(x) = [2^{(1-\alpha)x}
-1]/\gamma$. Plugging this into Eq.(\ref{I7}) the measure of
transmitted information will be
\begin{equation}
{\cal{I}}_\alpha({\cal{P}},\Im ) = \frac{1}{(1-\alpha)}\,
\mbox{log}_2 \left( \sum_{k=1}^{n}p^{\alpha}_k\right) \, .
\label{I80}
\end{equation}
\noindent The information measure (\ref{I80}) is usually called
the {\em generalized information measure} or {\em information
measure of order $\alpha$}, or simply {\em R{\'e}nyi's entropy}.
We will denote the explicit order of R{\'e}nyi's entropy as a
subscript in ${\cal{I}}(\ldots)$.

\vspace{3mm}

Although the foregoing {\em operational} (pragmatic) way of
arguing is quite robust, some readers may find more justifiable to
see R{\'e}nyi's entropy properly axiomatized. Actually, the
Shannon entropy was firstly axiomatized by Shannon\cite{Sh1} and
then later some axioms were weakened (or substituted) by
Fadeev\cite{Fa1}, Khinchin\cite{Kh1} and several other
authors\cite{CHau1}. The R{\'e}nyi entropy was axiomatized by
R{\'e}nyi himself\cite{Re1,Re2} and afterwards sharpened by
Dar\'{o}tzy\cite{Dar1} and others\cite{Oth2}. In further
considerations we will find, however, useful to use a slightly
different set of axioms than those utilized
in\cite{Re1,Re2,Dar1,Oth2}. In fact, in Appendix B we show that
the information measures (\ref{I79}) and (\ref{I80}) can be
characterized by the following axioms:
\vspace{3mm}
%
\begin{enumerate}
\item For a given integer $n$ and given ${\cal{P}} =
\{ p_1, p_2, \ldots , p_n\}$ ($p_k \geq 0, \sum_k^n p_k =1$),
${\cal{I}}({\cal{P}})$ is a continuous with respect to all its
arguments.

\item
For a given integer $n$, ${\cal{I}}(p_1, p_2, \ldots , p_n)$ takes
its largest value for $p_k = 1/n$ ($k=1,2, \ldots, n$) with the
normalization ${\cal{I}}\left( \frac{1}{2}, \frac{1}{2}\right)
=1$.

\item For a given $\alpha\in\mathbb{R}$;
${\mathcal{I}}({\mathcal{A}}\cap{\mathcal{B}}) =
{\mathcal{I}}({\mathcal{A}}) +
{\mathcal{I}}({\mathcal{B}}|{\mathcal{A}})$ with\\
\\
{\hspace{1cm}} ${\mathcal{I}}({\mathcal{B}}|{\mathcal{A}}) =
f^{-1} \left(\sum_k \varrho_k(\alpha)
f({\mathcal{I}}({\mathcal{B}}|{\mathcal{A}}={\mathcal{A}}_k))
\right)$,\\
\\
and $\varrho_k(\alpha) = (p_k)^{\alpha}/\sum_k (p_k)^{\alpha}$\,
(distribution ${\mathcal{P}}$ corresponds to the experiment
${\mathcal{A}})$.

%
\item $f$ is invertible and positive in $[0, \infty)$.

\item ${\cal{I}}(p_1,p_2, \ldots , p_n, 0 ) =
{\cal{I}}(p_1,p_2, \ldots , p_n)$, i.e., adding an event of
probability zero (impossible event) we do not gain any new
information.
\end{enumerate}

\subsection{Some observations about R\'{e}nyi's entropy}

%
Before going further let us observe some key characteristics of
Renyi's entropy which will prove essential in following sections.

\vspace{3mm}

(a)~${\mathcal{I}}_{\alpha}({\mathcal{B}}|{\mathcal{A}})$
appearing in the axiom 3 can be viewed as conditional information.
In fact, in Appendix~C we show that
${\mathcal{I}}_{\alpha}({\mathcal{B}}|{\mathcal{A}}) =0$ iff
outcome ${\mathcal{A}}$ uniquely determines outcome
${\mathcal{B}}$. We also show that when ${\mathcal{A}}$ and
${\mathcal{B}}$ are independent then
${\mathcal{I}}_{\alpha}({\mathcal{B}}|{\mathcal{A}}) ={
\mathcal{I}}_{\alpha}({\mathcal{B}})$ and hence
${\mathcal{I}}_{\alpha}({\mathcal{A}}\cap {\mathcal{B}}) =
{\mathcal{I}}_{\alpha}({\mathcal{A}}) +
{\mathcal{I}}_{\alpha}({\mathcal{B}})$, as expected. Alas the
reverse implication (i.e.,
${\mathcal{I}}_{\alpha}({\mathcal{B}}|{\mathcal{A}}) ={
\mathcal{I}}_{\alpha}({\mathcal{B}})$ $\Rightarrow$
${\mathcal{A}}$ and ${\mathcal{B}}$ are independent) generally
holds only when ${\mathcal{B}}$ has uniform distribution.

\vspace{3mm}

(b)~It is interesting to note that we can write (with a bit of
hindsight) in the axiom 3
\begin{displaymath} {\mathcal{I}}_{\alpha}({\mathcal{B}}|{\mathcal{A}}) =
f^{-1}\left(\sum_k
\varrho_k(\alpha)f({\mathcal{I}}_{\alpha}({\mathcal{B}}|{\mathcal{A}}=
{\mathcal{A}}_k ))\right)\, .
\end{displaymath}
\noi Similarly, we can write Eq.(\ref{I7}) as
\begin{displaymath}
{\mathcal{I}}({\mathcal{P}}) = f^{-1}\left(
\sum_k\varrho_k(1)f({\mathcal{I}}_1({\mathcal{A}}={\mathcal{A}}_k))\right)\,
. \end{displaymath}
\noi This indicates that when the constituent information of order
$\alpha$ enter a mean value calculation they must be weighted by
$\varrho_k(\alpha)$'s and not $p_k$'s, and this should hold true
whatever the Kolmogorov--Nagumo function is. The former result may
be generalized in the following way: Whenever outcomes of a
measurement carry an information of order $\alpha$ they must be
weighted with $\varrho_k(\alpha)$. When outcomes actually carry
information of order $\alpha$ will be discussed in Section IV~B.

\vspace{3mm}

(c)~Another important property of R{\'e}nyi's entropy is its
concavity for $\alpha < 1$ (for $\alpha
> 1$ R{\'e}nyi's entropy is not purely convex nor purely concave).
This a simple consequence of the fact that both $\log_2(x)$ and
$x^{\alpha}$ ($\alpha <1$) are concave functions (while
$x^{\alpha}$ is  convex for $\alpha >1$).

\vspace{3mm}

(d)~A notable point which we will use in Section IV is that
${\cal{I}}_{\alpha}$ is a monotonous decreasing function of
$\alpha$. This might be seen from the inequality
\begin{eqnarray}
\frac{d{\cal{I}}_{\alpha} }{d \alpha}
&=&  \ \frac{1}{(1-\alpha)^2}\left\{ - \log_2 \left\langle
{\cal{P}}^{1-\alpha} \right\rangle_{\alpha}  + \left\langle \log_2
{\cal{P}}^{1-\alpha}  \right\rangle_{\alpha}
\right\} \nonumber \\
& \leq & \ 0\, . \label{mon1}
\end{eqnarray}
\noi Here the expectation value $\langle \ldots \rangle_{\alpha} $
is defined with respect to the distribution $\varrho_k(\alpha)$.
The last
line of (\ref{mon1}) is due to Jensen's inequality and due to
concavity of $\log_2 (x)$.
%
%
Note that $d{\cal{I}}_{\alpha}/d\alpha =0$ only when the  Jensen
inequality used in the derivation (\ref{mon1}) is an equality.
This happen iff ${\cal{P}} = const.$ (see e.g.,\cite{Hardy1}), or
in other words when ${\cal{P}}$ is uniform. Consequently either
${\mathcal{I}}_{\alpha}$ is a strictly monotonous decreasing
function of $\alpha$ or all ${\mathcal{I}}_{\alpha}$ are
identical. One never finds, for example,
${\cal{I}}_{\alpha_1} < {\cal{I}}_{\alpha_2} =
{\cal{I}}_{\alpha_3}$ for $\alpha_1 > \alpha_2 > \alpha_3$.

\subsection{R{\'e}nyi's entropy and Shannon's entropy}

Now we turn to the investigation of the information measure of
order 1. An important element in this discussion is the fact that
$\mathcal{I}_\alpha$ is analytic in $\alpha =1$. This can be seen
by continuing the index $\alpha$ into the complex plane and
inspecting the behavior of $\log_2\left(\sum_{k=1}^n p^z_k\right)$
for $z \in\mathbb{C}$. The former is analytic provided that
$\sum_{k=1}^n p^z_k$ is not laying on the negative real axis. Let
us now consider the situation where $z = 1 + r \ e^{i\varphi}$
(i.e., we draw a circle with the radius $r$ centered at $z=1$).
Thus $\log_2\left(\sum_{k=1}^n p^z_k\right)$ is analytic
throughout the entire complex plane except the regions where the
following two conditions hold
\begin{eqnarray}
&& \sum_{k=1}^n \sin\left( r \sin\varphi \ln(p_k) \right) = 0\, ,\nonumber \\
&& \sum_{k=1}^n p_k^{r \cos\varphi +1} \cos\left(r \sin\varphi
\ln(p_k) \right) \leq 0\, . \label{anal1}
\end{eqnarray}
\noindent Let us put $r <
|\pi/(2\ln(p_k)_{\scriptsize\mbox{min}})|$. Then evidently for
such $r$'s the conditions (\ref{anal1}) cannot be fulfilled
together and we are safely in the analyticity region. Consider the
contour integral
\begin{equation}
I = \oint dz \ \frac{\log_2\left(\sum_{k=1}^n p^z_k\right)}{1-z}=
\oint dz \ {\mathcal{I}}_z({\mathcal{P}})\, , \label{anal2}
\end{equation}
\noindent around a contour $z = 1 + r \ e^{i \varphi}, \; \varphi
\in [0, 2\pi)$. The residue theorem assures then that
(\ref{anal2}) vanishes and as a result Renyi'e entropy is analytic
everywhere inside the contour (so also at $z = 1$). This shows
that the singularity of ${\mathcal{I}}_{\alpha}({\mathcal{P}})$ at
$\alpha =1$ is only spurious and, in fact, Renyi's entropy is
differentiable at $\alpha =1$ to all orders. Using the Cauchy
formula we can directly write
\begin{eqnarray}
{\mathcal{I}}_1({\mathcal{P}}) &=& \frac{i}{2\pi} \oint dz \
\frac{\log_2\left(\sum_{k=1}^n p^z_k\right)}{(1-z)(1-z)}\nonumber
\\
&=& \frac{1}{2\pi i}\oint dz \ \left(\frac{d}{dz}\frac{1}{(z-1)}
\right) \log_2\left(\sum_{k=1}^n p^z_k \right)\nonumber \\
&=& \frac{i}{2 \pi } \oint dz \  \frac{\sum_{k=1}^n p_k^z
\log_2(p_k)}{(z-1) \sum_{k=1}^n p_k^z} \nonumber \\
&=& - \sum_{k=1}^n p_k \log_2(p_k) =
{\mathcal{H}}({\mathcal{P}})\, , \label{anal3}
\end{eqnarray}
\noindent where the contour of integration is the same as in the
case (\ref{anal2}). It is usually argued that it is a matter of
modification of one of Shannon's axioms to get R{\'e}nyi's
entropy. We, however, do not intend to follow this path simply
because the Shannon entropy, as we have just seen, can be uniquely
determined from the behavior of (analytically continued)
R{\'e}nyi's entropy in the vicinity of $z=1$. In fact, we even do
not need to be in the vicinity because the circle used in the
contour integral (\ref{anal3}) can be analytically continued to
any curve which lies in the $1$st and $4$th quadrant and which
encircles the point $z=1$. View which we intend to advocate here
is that the Shannon entropy is not a special information measure
deserving separate axiomatization but a member of a wide class of
entropies embraced by a single unifying axiomatics.

\vspace{3mm}
An important consequence of the fact that ${\mathcal{I}}_{\alpha}$
is a monotonous decreasing function of $\alpha$ is embodied in the
following two inequalities
\begin{eqnarray}
&&{\cal{H}} < {\cal{I}}_{\alpha} < \log_2 n\, ,\;\;\;\; 0 < \alpha < 1\,
,\label{2.22} \\
&&{\cal{I}}_{\alpha} <{\cal{H}}< \log_2 n\, ,\;\;\;\; \alpha > 1\,
. \label{2.23}
\end{eqnarray}
\noindent Inequality (\ref{2.23}) shows that ${\cal{H}}$
represents an upper bound for all R\'{e}nyi entropies with $\alpha
>1$. This finding will play an important r\^{o}le in the
reconstruction theorem in Section~\ref{rec}.

\subsection{R{\'e}nyi's entropy and THC entropy }

Due to an increasing interest in  long--range correlated systems
and  non--equilibrium phenomena there has been currently much
discussed the, so called, Tsallis (or non--extensive) entropy.
Although firstly introduced by Havrda and Charvat in the
cybernetics theory context\cite{HaCh} it was Tsallis\cite{Ts1} who
exploited its non--extensive features and placed it in a physical
setting. THC entropy reads
\begin{equation}
{\cal{S}}_{\alpha} =
\frac{1}{(1-\alpha)}\left[\sum_{k=1}^{n}(p_k)^{\alpha} -1\right]
\, , \;\;\;\; \alpha > 0 \, .
\label{Thc2}
\end{equation}
\noindent The most important properties of THC entropy can be
easily read out of (\ref{Thc2}). For instance, employing Jensen's
inequality we have for $\alpha > 1$ that $\sum_k p_k^{\alpha} \leq
1 $ (while for $0<\alpha<1$ the reverse inequality holds) and
hence ${\cal{S}}_{\alpha}$ is non--negative. Similarly, choosing
any pair of distributions ${\cal{P}}$ and ${\cal{Q}}$, and a real
number $0 \leq \lambda \leq 1$ we have
\begin{eqnarray}
{\cal{S}}_{\alpha}(\lambda {\cal{P}} + (1-\lambda) {\cal{Q}})
 = \, \lambda \,
{\cal{S}}_{\alpha}({\mathcal{P}}) + (1-\lambda)\,
{\cal{S}}_{\alpha}({\mathcal{Q}})\, , \label{Min1}
\end{eqnarray}
\noindent and so THC entropy is a concave function of its
probability distribution. Eq.(\ref{Min1}) results from Jensen's
inequality a concavity of $x^{\alpha}/(1-\alpha)$. In addition, by
rule of l'Hospital we get that
\begin{equation}
\lim_{\alpha \rightarrow 1}{\cal{S}}_{\alpha} = \lim_{\alpha
\rightarrow 1}{\cal{I}}_{\alpha} = {\cal{H}} \, . \label{ren33}
\end{equation}
\noindent Thus in the $\alpha \rightarrow 1$ limit THC entropy
reduces to Shannon's entropy.

\vspace{3mm}

Perhaps the most distinguished feature of THC entropy is the so
called pseudo--additivity\cite{Ts1,Abe}
\begin{displaymath}
{\cal{S}}_{\alpha}({\cal{A}} \cap {\cal{B}}) =
{\cal{S}}_{\alpha}({\cal{A}}) +
{\cal{S}}_{\alpha}({\cal{B}}|{\cal{A}}) + (1-\alpha)
{\cal{S}}_{\alpha}({\cal{A}}){\cal{S}}_{\alpha}({\cal{B}}|{\cal{A}})
 \, ,
\end{displaymath}
\noi for two experiments ${\cal{A}}$ and ${\cal{B}}$,
${\cal{S}}_{\alpha}({\cal{B}}|{\cal{A}})$ represents here the
conditional THC entropy. Remarkable, albeit not yet understood
aspect of the pseudo--additivity is that in the case of
independent experiments THC entropy is not additive. Interested
reader may find further discussion of THC entropy, for instance,
in Ref.\cite{Ts2}.

\vspace{3mm}

Now we turn to the problem of finding the connection between
R{\'e}nyi's and THC entropy.
%
%
%
To this end we utilize the identity
\begin{eqnarray}
{\cal{I}}_\alpha &=& \frac{1}{(1-\alpha)}\log_2
\left[(1-\alpha){\cal{S}}_{\alpha} + 1\right]\nonumber \\
&=& \frac{1}{k} \int_{0}^{{\mathcal{S}}_{\alpha}} dx \ \frac{1}{1+
x (1-\alpha) }\, . \label{THCR1}
\end{eqnarray}
\noindent Here $k = \ln2$ is the scale factor. For
$|(1-\alpha){\mathcal{S}}_{\alpha}| < 1$ we may expand the
integrand in (\ref{THCR1}). In such a case the (geometric) series
is absolutely convergent and we can integrate it term by term:
\begin{equation}
{\mathcal{I}}_{\alpha} = \frac{1}{k}{\mathcal{S}}_{\alpha} -
\frac{1}{2k} (1-\alpha){\mathcal{S}}^2_{\alpha} +
{\cal{O}}\left[(1-\alpha)^2{\mathcal{S}}^3_{\alpha}\right] \, .
\label{THCR2}\end{equation}
\noi So apart from an unimportant factor $k$ (which just sets the
scale for entropy units) we see that ${\cal{I}}_\alpha \approx
{\cal{S}}_{\alpha}$, provided
\begin{equation}
|(1-\alpha) {\mathcal{S}}_{\alpha}| =  \left| \sum_l^n
(p_l)^{\alpha} -1 \right| \ll 1 \, . \label{in1}
\end{equation}
\noindent It should be understood that the expansion (\ref{THCR2})
is not necessarily the expansion in $(1-\alpha)$. In fact,
condition (\ref{in1}) may be fulfilled in numerous ways.
Obviously, for $\alpha \approx 1$ the inequality (\ref{in1}) is
trivially satisfied. This should be expected because both
${\cal{I}}_\alpha$ and ${\cal{S}}_{\alpha}$ tend to the same limit
value at $\alpha \approx 1$. Thus the actual error estimate in
this instance can be written as
\begin{equation}
{\cal{I}}_\alpha = \frac{1}{k} {\cal{S}}_{\alpha} + {\cal{O}}\left( (\alpha -1) \,
{\cal{H}}^2 \right) \, ,
\end{equation}
\noindent and so the true inaccuracy in dealing with
${\cal{S}}_{\alpha}$ and not ${\cal{I}}_\alpha$ is of order
$(\alpha -1)$. There is, however, possible to pinpoint other very
important classes of systems with $\alpha \not\approx 1$ still
obeying (\ref{in1}). Clearly, various improved estimates can be
devised if some additional assumptions are made about the system.
One particularly important case which is pertinent to $\alpha <1$
region, namely the case of large deviations will be briefly
discussed now.

\vspace{3mm}

%
%
%
Systems with large deviations prove fruitful in many areas of
physics and mathematics ranging from fluid dynamics and weather
forecast to population breeding. To proceed we will appeal to
Lo\'{e}ve (or basic) inequality of probability theory\cite{Lo1}.
Let $X$ be an arbitrary random variable and let $g$ be an even
function on $\mathbb{R}$ and non--decreasing on $[0,\infty)$. Then
for $\forall \, a \geq 0$
\begin{equation}
\langle g(X)\rangle - g(a)\, \leq \, \sup g(X)\, P[\ |X| \geq a]\,
. \label{lo1}
\end{equation}
\noindent
Upon taking the distribution $\varrho(q)=\{(p_k)^{q}/\sum_k
(p_k)^q\},$ $q\in [0,1]$ and $g(x) = |x|^{\alpha - q},$ $\alpha\in
[0,1]$ we get from (\ref{lo1})
\begin{equation}
\left\langle |X|^{\alpha-q} \right\rangle_q -  a^{\alpha-q}\, \leq
\, \sup(|X|^{\alpha -q})\,P[\ |X| \geq a ]\, . \label{lo2}
\end{equation}
\noindent Here $\langle \ldots \rangle_q$ is the mean with respect
to $\varrho(q)$. We can now set $|X| = {\mathcal{P}} =\{ p_k \}$
and fix $q$ so to fulfill $\alpha > q$. Taking
\begin{equation}
\frac{1}{a} = \left(\sum_k^n (p_k)^{q}\right)^{\!\!1/(\alpha -q)}
\!\!\!\!\!\!\!\equiv \, Z(q)^{1/(\alpha -q)}\, ,
\end{equation}
\noi we obtain the probability theory variant of (\ref{in1}),
namely
\begin{eqnarray}
\sum_k^n (p_k)^{\alpha} - 1 \, &\leq& \,
\sup({\mathcal{P}}^{\alpha
-q })\ P[{\mathcal{P}} \geq a] \ Z(q)\nonumber \\
&\leq& \, P[{\mathcal{P}} \geq a] \ Z(q) \, . \label{lo3}
\end{eqnarray}
\noindent To proceed we realize that for $q \in [0,1]$ we have
$1\leq Z(q)\leq n^{1-q}$ and hence
\begin{equation}
1\geq a \geq \left( \frac{1}{n} \right)^{\!\!(1-q)/(\alpha -q) }\,
.
\end{equation}
\noi Note particularly that $(1-q)/(\alpha -q)\! > \! 1$. Thus if
for most of  $i$'s the inequality $p_i \leq (1/n)^{(1-q)/(\alpha
-q)}$ holds (rare events) then $P[{\mathcal{P}} \geq a]$ of
(\ref{lo3}) can be made arbitrarily small\footnote{Of course, due
to normalization condition $\sum_i^n p_i =1$, $P[{\mathcal{P}}
\geq a]$ cannot be zero since there must be always a very small
probability for large (i.e., $> 1/n$) $p_i$'s. Hence name {\em
large deviations}.}. Besides, because $Z(q)$ is bounded by
$n^{1-q}$ irrespective of a particular choice of ${\mathcal{P}}$
and $\alpha$ we may use this freedom to fix RHS of (\ref{lo3}) to
be very small. So for example when most $p_i \approx 1/n^2$ then
the choice $q = 1/2$ and $\alpha = 3/4$ assure that $Z(q)\leq
{\sqrt{n}}$ while $P[{\mathcal{P}}\geq a] \approx 1/n$ and hence
RHS of (\ref{lo3}) is smaller than $1/\sqrt{n}$. It should be
recognized that in this case the inequality (\ref{in1}) holds not
because $\alpha \! \rightarrow \! 1$ but because $n$ is large.

\vspace{3mm}

It is interesting to consider now the situation when $|(1-\alpha)
{\mathcal{S}}_{\alpha}|>1$. Such a case is undoubtedly more
intriguing than the previous one as it represents a wider class of
physically relevant situations. Let us start first with
the situation $|(1-\alpha){\mathcal{S}}_{\alpha}|\approx 1$.
There are two cases of interest here. The case when
$(1-\alpha){\mathcal{S}}_{\alpha}\approx 1$ is the simpler one. Here
$\alpha < 1$ due to positivity of ${\mathcal{S}}_{\alpha}$ and we may
rewrite (\ref{THCR1}) as
\begin{eqnarray}
k{\mathcal{I}}_{\alpha} &=& \left( \int_{0}^{1/(1-\alpha)} +
\int_{1/(1-\alpha)}^{{\mathcal{S}}_{\alpha}}\right) dx \
\frac{1}{1+x(1-\alpha)}\nonumber \\
&&\  \nonumber \\
&=& \frac{k}{(1-\alpha)} + \frac{{\mathcal{S}}_{\alpha} -
1/(1-\alpha)}{2}\nonumber\\
&&\nonumber \\ && + \, {\mathcal{O}}\left(\frac{
[(1-\alpha){\mathcal{S}}_{\alpha} -1]^2}{(1-\alpha)}
\right)\nonumber \\
&&\  \nonumber \\
&\approx& \frac{{\mathcal{S}}_{\alpha}}{2} +
\frac{1}{(1-\alpha)}\left(k-\frac{1}{2}\right) \, . \label{a1}
\end{eqnarray}
\noi On the other hand, the case when
$(1-\alpha){\mathcal{S}}_{\alpha} \approx -1$ is very important as
it corresponds to the large $\alpha$ limit. Since for high
$\alpha$,  ${\mathcal{S}}_{\alpha}$ asymptotically approaches
$\zeta = [(p_k)_{\mbox{\scriptsize{max}}}^{\alpha} -1]
/(1-\alpha)$ from above we can write
\begin{eqnarray}
k{\mathcal{I}}_{\alpha} &=& \left( \int_{0}^{\zeta}
+ \int_{\zeta}^{{\mathcal{S}}_{\alpha}}\right) dx \
\frac{1}{1+x(1-\alpha)}\nonumber \\
&&\  \nonumber \\
&=& \frac{\alpha \ln (p_k)_{\mbox{\scriptsize{max}}}}{(1-\alpha)} +
\frac{{\mathcal{S}}_{\alpha}(1-\alpha) + (1-(p_k)_{\mbox{\scriptsize{max}}}^{\alpha})
}{(1-\alpha)(p_k)_{\mbox{\scriptsize{max}}}^{\alpha}}\nonumber \\
&&\  \nonumber \\
&& + \, {\mathcal{O}}\left( [{\mathcal{I}}_{\alpha}
+\log_2(p_k)_{\mbox{\scriptsize{max}}}]^2
\right)\nonumber \\
&&\  \nonumber \\
&\approx& \frac{{\mathcal{S}}_{\alpha}
}{(p_k)_{\mbox{\scriptsize{max}}}^{\alpha}} +
\frac{\left(1 + (p_k)_{\mbox{\scriptsize{max}}}^{\alpha}[\alpha
\ln(p_k)_{\mbox{\scriptsize{max}}} - 1]
\right)}{(1-\alpha)(p_k)_{\mbox{\scriptsize{max}}}^{\alpha}}
\, .\nonumber \\
\label{a2}
\end{eqnarray}
\noi In both previous cases we have seen that the leading orders
yielded a linear relationship between R\'{e}nyi's and THC entropy.
As already recognized by Schr\"{o}dinger\cite{Sch2}, statistical
entropy is defined up to a linear transformation. This, in turn,
one could view as a conceptual backing for THC entropy in the
respective situations. Ones pleasure is short--lived, however,
when one starts to consider the case
$(1-\alpha){\mathcal{S}}_{\alpha} \gg 1$. This corresponds, for
example, to the situation when $\alpha \rightarrow 0$. Writing
(\ref{THCR1}) as
\begin{eqnarray}
k{\mathcal{I}}_{\alpha} &=& \left( \int_{0}^{1/(1-\alpha)} +
\int_{1/(1-\alpha)}^{{\mathcal{S}}_{\alpha}}\right) dx \
\frac{1}{1+x(1-\alpha)}\nonumber \\
&&\  \nonumber \\
&= & \ \frac{k}{(1-\alpha)} + \sum_{n=0}^{\infty} (-1)^n
\int_{1/(1-\alpha)}^{{\mathcal{S}}_{\alpha}} dx \ \left(
\frac{1}{x(1-\alpha)} \right)^{\!\! n+1}
\nonumber \\
&&\  \nonumber \\
&\approx & \
\frac{\ln({\mathcal{S}}_{\alpha}(1-\alpha))}{(1-\alpha)} +
\frac{1}{{\mathcal{S}}_{\alpha}(1-\alpha)^2} \, , \label{a3}
\end{eqnarray}
\noi we see that there is a logarithmic singularity at large
${\mathcal{S}}_{\alpha}$. Hence, no linear mapping between RHC and
R\'{e}nyi's entropy exists in this region. One may thus expect
that for $(1-\alpha){\mathcal{S}}_{\alpha} \gg 1$ both entropies
have qualitatively different behavior and the conceptual grounding
for THC entropy must be sought out of the scope of information
theory.

\vspace{3mm}

Let us add two more comments. It is often argued that concavity of
THC entropy with respect to probability distribution makes it
better suited, say, for thermodynamic considerations. It is,
however, concavity with respect to extensive variables rather than
probability distribution which ensures stability of thermodynamic
equilibrium\cite{Rb1}. The first does not necessarily implies the
second. Needless to say that there is no general concavity
requirement for entropy in non--equilibrium systems. Secondly,
from Eq.(\ref{THCR1}) we see that THC entropy and R{\'e}nyi's
entropy are monotonic functions of each other and, as a result,
both must be maximized by the same probability distribution.
However, while R{\'e}nyi's entropy is additive, THC entropy is
not, so that it appears that the additivity property is not
important for entropies required for maximization purposes.

\section{R{\'e}nyi's entropy of continuous probability
distributions}

While in the previous section we dealt with the R{\'{e}}nyi's
entropy of discrete probability distributions we will now discuss
the corresponding continuous counterpart. We shall see that in the
latter case a host of new properties will emerge.
As a byproduct we get a consistent extension of THC entropy for
continuous distributions.

\vspace{3mm}

Let us first assume that ${\cal{F}}(x)$ is an arbitrary {\em
continuous}, positive density function (PDF) defined, say, in the
interval $[0,1]$. By defining the integrated probability
\begin{displaymath}
p_{nk} = \int_{k/n}^{(k+1)/n}dx \,{\cal{F}} (x)\, ; \,\,\,\,\,\, k =
0,1, \ldots , n-1 \, ,
\end{displaymath}
\noindent we generate the discrete distribution ${\mathcal{P}}_n =
\{p_{nk}\} $. It might be then shown\cite{Re1,Re2} that
\begin{eqnarray}
{\mathcal{I}}_{\alpha}({\cal{F}}) &\equiv& \lim_{n \rightarrow
\infty}({\mathcal{I}}_{\alpha}({\mathcal{P}}_n) - \log_2 n
)\nonumber \\
&=& \frac{1}{1-\alpha}\, \log_2 \left( \int_0^1 dx\, {\cal{F}}^{\alpha}(x)
\right)\, , \label{ren1}
\end{eqnarray}
\noindent provided that $\int_0^1 dx\, {\cal{F}}^{\alpha}(x)$
exists\footnote{For $0< \alpha < 1$ this is always the case as
$\sum_k (p_{nk})^{\alpha} \le n^{1-\alpha} \, \Rightarrow \,
\int_0^1 dx\, {\cal{F}}^{\alpha}(x) \le 1 $.}. Here $\log_2 n$
must be subtracted to ensure a correct measure in the integral.
Defining the uniform distribution ${\cal{E}}_n = \left\{
\frac{1}{n}, \ldots, \frac{1}{n} \right\}$ then $\log_2 n =
{\cal{I}}_{\alpha}({\cal{E}}_n)$. From this we may interpret
$-{\mathcal{I}}_{\alpha}({\cal{F}}) \sim
{\cal{I}}_{\alpha}({\cal{E}}_n)
 - {\cal{I}}_{\alpha}({\cal{P}}_n) $ as the gain of information
obtained by replacing the uniform distribution ${\cal{E}}_n$
(having maximal uncertainty) by distribution ${\cal{P}}_n$ or, in
other words, $-{\mathcal{I}}_{\alpha}({\cal{F}})$ represents the
decrease of uncertainty when ${\cal{E}}_n$ is replaced by
${\cal{P}}_n$. In the case of Shannon's entropy the quantity
$-{\cal{H}}({\cal{F}})$ is usually called the informative content
or ``negentropy'' and states how much uncertainty is still left
unresolved after a measurement (for discussion see
e.g.,{\cite{Sz1,Br1}).

\vspace{3mm}

Relation (\ref{ren1}) can be viewed as a {\em renormalized}
R\'{e}nyi's information content. This may be understood from the
asymptotic expansion of ${\cal{I}}_{\alpha}({\cal{P}}_n)$, namely
\begin{equation}
{\cal{I}}_{\alpha}({\cal{P}}_n) = \mbox{divergent in}\ n +
\mbox{finite} + o(1)\, ,
\end{equation}
\noi the $o(1)$ symbol means that the residual error tends to $0$
for $n \rightarrow \infty$. The $finite$ part ($=
{\mathcal{I}}_{\alpha}({\cal{F}}) $) is fixed by requirement (or
by renormalization prescription) that it should fulfill the
postulate of additivity in order to be identifiable with an
information measure. Incidentally, the latter uniquely identifies
the divergent part as $\log_2 n$. The above renormalization
procedure is somehow analogous to that in quantum field theory
where one renormalizes energy by subtracting the ground state
contribution. It should be, however, noted that the information
$\log_2 n$ is usually greater than
${\mathcal{I}}_{\alpha}({\mathcal{P}}_{nk})$ and consequently
${\mathcal{I}}_{\alpha}({\cal{F}})$ is not positive. The former
should be contrasted with the discrete case where
${\cal{I}}_{\alpha}$ is by construction non--negative.

\vspace{3mm}

Extension of (\ref{ren1}) into $d$--dimensional situations is
straightforward. Having a $d$--dimensional random variable (i.e.,
experiment) ${\cal{A}}^{\scriptscriptstyle (d)}$ we can discretize
it in the following way; ${\cal{A}}_n^{\scriptscriptstyle (d)} =
\left( \frac{[n{\cal{A}}_1]}{n}, \frac{[n{\cal{A}}_2]}{n}, \ldots,
\frac{[n{\cal{A}}_d]}{n} \right)$ where $[\ldots]$ denotes
integral part. This divides the $d$--dimensional volume $V$ of the
outcome (or sample) space into boxes labelled by an index $k$
which runs from $1$ up to $[Vn^d]$. The size of the $k$th box is
$l = 1/n$ and its probability distribution
${\cal{P}}_n^{\scriptscriptstyle (d)} =
\{p_{nk}^{\scriptscriptstyle (d)}
 \}$ is generated via prescription
\begin{displaymath}
 p_{nk}^{\scriptscriptstyle (d)} = \int_{k{\mbox{{\footnotesize
th box}}}} d^d{\bf x}\,
{\cal{F}}({\bf x}) \,; \,\,\,\, k = 1, 2, \ldots, [Vn^d]  \, .
\end{displaymath}
\noi It can be shown then (see e.g.,\cite{Re1} and Appendix D)
that
\begin{eqnarray}
{\cal{I}}_{\alpha}^{\scriptscriptstyle (d)}({\cal{F}})
&\equiv& \lim_{n \rightarrow \infty}
({\cal{I}}_{\alpha}({\cal{P}}_n^{\scriptscriptstyle (d)})
- d \log_2 n ) \nonumber \\
& = & \frac{1}{(1-\alpha)} \, \log_2 \left(\int_V d^d {\bf{x}} \,
{\cal{F}}^{\alpha}({\bf{x}}) \right)\, ,
\label{ren2}
\end{eqnarray}
\noi provided that $\int_V d^d {\bf{x}} \,
{\cal{F}}^{\alpha}({\bf{x}})$ exists.

\vspace{3mm}

Question now stands whether we get unique
${\cal{I}}_{\alpha}^{\scriptscriptstyle (d)}({\cal{F}})$ by
mimicking the previous recipe, i.e., performing the asymptotic
expansion of ${\cal{I}}_{\alpha}({\cal{P}}_n^{\scriptscriptstyle
(d)})$ and pinpointing the correct finite part by the
renormalization condition - additivity of information. In the
non--unit volume, however, one more fixing condition is required.
To see that we define the uniform distribution
${\cal{E}}^{\scriptscriptstyle (d)}_n = \left\{ \frac{1}{V_n n^d},
\ldots, \frac{1}{V_n n^d} \right\}$ with $V_n \equiv \frac{[V
n]}{n} \stackrel{n \rightarrow \infty}{\longrightarrow} V$.
R{\'e}nyi's entropy then reads
\begin{displaymath}
{\cal{I}}_{\alpha}({\cal{E}}^{\scriptscriptstyle (d)}_n) = \log_2 V_n
+ d\log_2 n \, ,
\end{displaymath}
\noi and so
\begin{eqnarray}
\tilde{{\cal{I}}}_{\alpha}^{\scriptscriptstyle (d)}({\cal{F}})
&\equiv& \lim_{n \rightarrow \infty}
({\cal{I}}_{\alpha}({\cal{P}}_n^{\scriptscriptstyle (d)})
- {\cal{I}}_{\alpha}({\cal{E}}_n^{\scriptscriptstyle (d)}) ) \nonumber \\
& = & \frac{1}{(1-\alpha)} \, \log_2 \left(\frac{\int_V d^d
{\bf{x}} \, {\cal{F}}^{\alpha}({\bf{x}})}{\int_V d^d {\bf{x}} \,
1/V^{\alpha}} \right) \, .
\label{ren3}
\end{eqnarray}
\noi Alike in (\ref{ren2}) the RHS of (\ref{ren3}) represents the
finite part in
 the
asymptotic expansion of
${\cal{I}}_{\alpha}({\cal{P}}_n^{\scriptscriptstyle (d)})$, the
part which fulfils the additivity of information condition.
To ensure the uniqueness of R{\'e}nyi entropy in the case of
continuous distributions we must, in addition, fix the value of
the finite part at ${{\cal{F}}} = (1/V)$. It is then matter of
taste and/or a particular problem at hand which convention should
be used. In this paper we will use the renormalization
prescription where ${\cal{I}}_{\alpha}^{\scriptscriptstyle
(d)}(1/V)|_{finite} = \log_2 V$ (i.e., the one which implies
Eq.(\ref{ren2})). The latter merely means that we define
R\'{e}nyi'e entropy with PDF ${\mathcal{F}}$ as
\begin{equation}
{\cal{I}}_{\alpha}^{\scriptscriptstyle (d)}({\cal{F}}) \equiv
\lim_{n \rightarrow \infty}
({\cal{I}}_{\alpha}({\cal{P}}_n^{\scriptscriptstyle (d)})
-{\cal{I}}_{\alpha}({\cal{E}}_n^{\scriptscriptstyle (d)})|_{V=1}
)\, .
\label{ren4}
\end{equation}
\noi  In Section~\ref{seciv} we generalize results (\ref{ren3})
and (\ref{ren4}) into fractal and multifractal systems. A comment
is in order. It may be shown (see Appendix E) that the form
(\ref{ren3}) is, in fact, a better candidate for the information
measure than (\ref{ren2}) as it is an invariant under a
transformation of ${\cal{A}}^{\scriptscriptstyle (d)}$. However,
difference between (\ref{ren2}) and (\ref{ren3}) is often only a
constant which ensures that for the questions we address here it
is quite adequate to use the simpler form (\ref{ren2}). It should
be, however, clear that there are system of physical interest
where the ground--state entropy plays a central r\^{o}le (e.g.,
frustrated spin systems or quantum liquids). In such cases the
form (\ref{ren3}) is obligatory.

\vspace{3mm}

Let us now examine the implications of (\ref{ren1})--(\ref{ren3})
for THC entropy with continuous distributions. For this we will
use the convention introduced before Eq.(\ref{ren2}). Firstly,
from (\ref{THCR1}) and (\ref{ren2}) follows that
$[{\cal{I}}_{\alpha}({\cal{P}}_n) - d\log_2 n ]$ is finite at
large $n$ (provided  $\int_V d^d {\bf{x}} \,
{\cal{F}}^{\alpha}({\bf{x}})$ exists) and so
\begin{equation}
\frac{(1-\alpha){\cal{S}}_{\alpha}({\cal{P}}_n) + 1}{n^{d(1-\alpha)}}
= \int_V d^d {\bf{x}} \,
{\cal{F}}^{\alpha}({\bf{x}})\, + \, o(1)\, .
\end{equation}
\noi In order to obtain the correct THC entropy with PDF
${\cal{F}}$ it is conceptually simplest to follow the same route
as before, i.e., asymptotically  expand
${\cal{S}}_{\alpha}({\cal{P}}_n)/n^{d(1-\alpha)}$ and look for the
finite part which conforms to certain renormalization
prescription\footnote{It is indeed
${\cal{S}}_{\alpha}({\cal{P}}_n)/n^{d(1-\alpha)}$ rather than
${\cal{S}}_{\alpha}({\cal{P}}_n)$ which should be asymptotically
expanded. For instance, for $0 < \alpha <1$ the asymptotic
expansion of ${\cal{S}}_{\alpha}({\cal{P}}_n)$ would be $o(1)$ and
so the corresponding large $n$ limit would be trivial. It is not
difficult to see that it is only the fraction
${\cal{S}}_{\alpha}({\cal{P}}_n)/n^{d(1-\alpha)}$ which has a
senseful meaning in the large $n$ limit.}. Unlike the R{\'e}nyi
entropy case we do not have now any first principle
renormalization prescription ({\em \`{a} la} additivity of
information) which we could impose. As a matter of fact, one could
be tempted to use the THC pseudo--additivity condition to isolate
the proper finite part in the
${\cal{S}}_{\alpha}({\cal{P}}_n)/n^{d(1-\alpha)}$ expansion, but
such a renormalization condition would be clearly {\em ad hoc} as
there is no a priori reason to assume that the non--extensivity
condition obeys the same prescription in the continuous case. It
is fairly safer to follow the analogy with Eqs.(\ref{ren3}) and
(\ref{ren4}) demanding, for instance, the consistency for
$\alpha$'s in the complex vicinity of $\alpha =1$ (i.e., values at
which R\'{e}nyi and THC entropies coincide). If the consistency is
reached then the validity of the result can be analytically
continued to the whole domain of analyticity of
${\mathcal{S}}_{\alpha}$ - so particularly to $\alpha \in
\mathbb{R}^+$.

\vspace{3mm}

\noi Using the asymptotic expansions:
\begin{eqnarray}
\frac{{\cal{S}}_{\alpha}({\cal{P}}_n^{\scriptscriptstyle
(d)})}{n^{d(1-\alpha)}} &=& - \frac{1}{(1-\alpha)n^{d(1-\alpha)}}\nonumber \\
&+& \frac{1}{(1-\alpha)}\int_V d^d {\bf{x}} \,
{\cal{F}}^{\alpha}({\bf{x}}) + o(1)\, , \nonumber \\
\frac{{\cal{S}}_{\alpha}({\cal{E}}_n^{\scriptscriptstyle
(d)})}{n^{d(1-\alpha)}} &=& - \frac{1}{(1-\alpha)n^{d(1-\alpha)}}\nonumber \\
&+& \frac{1}{(1-\alpha)}\int_V d^d {\bf{x}} \, 1/V^{\alpha} +
o(1)\, ,
\end{eqnarray}
\noi we may immediately write
\begin{eqnarray}
\tilde{{\cal{S}}}_{\alpha}^{\scriptscriptstyle (d)}({\cal{F}})
&\equiv& \lim_{n \rightarrow \infty} \left(
\frac{{\cal{S}}_{\alpha}({\cal{P}}_n^{\scriptscriptstyle
(d)})}{n^{d(1-\alpha)}} - \frac{{\cal{S}}_{\alpha}
({\cal{E}}_n^{\scriptscriptstyle
(d)})}{n^{d(1-\alpha)}} \nonumber \right)\\
&=& \frac{1}{(1-\alpha)}\ \left(\int_V d^d {\bf{x}}
\,{\cal{F}}^{\alpha}({\bf{x}}) -1 \right) \nonumber \\
 &-& \frac{1}{(1-\alpha)} \ \left(\int_V d^d {\bf{x}} \, 1/V^{\alpha} -1
\right)\, ,  \nonumber \\
~ \nonumber \\
{\cal{S}}_{\alpha}^{\scriptscriptstyle (d)}({\cal{F}})
&\equiv& \lim_{n \rightarrow \infty} \left(
\frac{{\cal{S}}_{\alpha}({\cal{P}}_n^{\scriptscriptstyle
(d)})}{n^{d(1-\alpha)}}  - \frac{{\cal{S}}_{\alpha}
({\cal{E}}_n^{\scriptscriptstyle
(d)})|_{V=1}}{n^{d(1-\alpha)}}\right) \nonumber \\
&=& \frac{1}{(1-\alpha)}\
\left(\int_V d^d {\bf{x}}
\,{\cal{F}}^{\alpha}({\bf{x}}) -1 \right)\, .
\label{ren5}
\end{eqnarray}
\noi It is not difficult to check that for $|\alpha|\in
[1-\epsilon, 1+\epsilon]\, , \,\varepsilon\ll 1,$ (\ref{ren5}) is
consistent with (\ref{ren3}) and (\ref{ren4}).

\vspace{3mm}

Let us note at the end that from the asymptotic expansion of
${\cal{I}}_{\alpha}({\cal{P}}_n^{\scriptscriptstyle (d)})$ i.e.,
from
\begin{equation}
{\cal{I}}_{\alpha}({\cal{P}}_n^{\scriptscriptstyle (d)}) = d
\log_2 n + {\cal{I}}_{\alpha}^{\scriptscriptstyle
(d)}({\mathcal{F}}) + o(1)\, , \label{D01}
\end{equation}
\noi we find, in return, that the dimension $d$ is identified with
\begin{equation}
d(\alpha) = \lim_{n \rightarrow \infty} \frac{{\cal{I}}_{\alpha}(
{\cal{P}}_n^{\scriptscriptstyle (d)})}{\log_2 n}\, .
\label{D1}
\end{equation}
\noi For simple metric (outcome) spaces (like ${\mathbb{R}}^d$) we
will prove in the following section that $d(\alpha) = d$ for all
$\alpha$ and it coincides with the usual topological dimension.
This situation is however not generic. In the next section we
shall see what modifications should be done when (multi)fractal
systems are in question.

\section{R{\'e}nyi's parameter and (multi)fractal
dimension}\label{seciv}

Fractals, objects with a generally non--integer dimension
exhibiting the scaling property and property of self--similarity
have had a significant impact not only on mathematics but also on
such distinctive fields as physical chemistry, astrophysics,
physiology, and fluid mechanics. The key characteristic of
fractals is fractal dimension which is defined as follows:
Consider a set $M$  embedded in a $d$--dimensional space. Let us
cover the set with a mesh of $d$--dimensional cubes of size $l^d$
and let $N_l(M)$ is a minimal number of the cubes needed for the
covering. The fractal dimension (or similarity dimension) of $M$
is then defined as\cite{Man1,Fed1}
\begin{equation}
D = - \lim_{l \rightarrow 0} \frac{\ln N_l(M)}{\ln l}\, .
\label{fra1}
\end{equation}
\noi  In most cases of interest the fractal dimension (\ref{fra1})
coincides with the Hausdorff--Besicovich fractal dimension
used by Mandelbrot\cite{Man1}.

\vspace{3mm}

Multifractals, on the other hand, are related to the study of a
distribution of physical or other quantities on a generic support
(be it or not fractal) and thus provide a move from the geometry
of sets as such to geometric properties of distributions. An
intuitive picture about an inner structure of multifractals is
obtained by introducing the $f(a)$ spectrum\cite{Pa1,Kad1}.  To
elucidate the latter let us suppose that over some support
(usually a subset of a metric space) is distributed a probability
of a certain phenomenon, be it e.g., probability of electric
charge, magnetic momenta, hydrodynamic vorticity or mass. If we
cover the support with boxes of size $l$ and denote the integrated
probability in the $i$th box as $p_i$, we may define the local
scaling exponent $a_i$ by
\begin{equation}
p_i (l) \sim l^{a_i}\, .
\label{p1a}
\end{equation}
\noi where $a_i$ is called the Lipshitz--H\"{o}lder exponent. Here
and throughout the symbol $\sim$ indicates an asymptotic relation,
e.g., (\ref{p1a}) should read:
\begin{displaymath}
a_i = \lim_{l \rightarrow 0} \frac{\ln\ p_i(l)}{\ln l}\, .
\end{displaymath}
\noi The proportionality constant (say $c(a_i)$) in (\ref{p1a})
can be weakly dependent on $l$.   By ``weakly" we mean that
\begin{displaymath}
\lim_{l\rightarrow 0} \frac{\ln c(a_i,l)}{\ln l} = 0\, .
\end{displaymath}
\noi  Note that PDF of each of small pieces is
\begin{equation}
\rho_i = \frac{p_i}{l^d}\sim l^{a_i -d}\, ,
\label{p1aa}\end{equation}
\noi and so $a_i$ controls the singularity of $\rho_i$. Inasmuch
$a_i$ is also known as the singularity exponent.

\vspace{3mm}

Counting number of boxes $dN(a)$ where $p_i$ has singularity
exponent between $a$ and $a + da$, then $f(a)$ defines the fractal
dimension of the set of boxes with the singularity exponent $a$ by
\begin{equation}
dN(a) \sim l^{-f(a)} da\, .
\label{N1a}
\end{equation}
\noi Here $f(a)$ is called singularity spectrum. Multifractal can
be then viewed as the ensemble of intertwined (uni)fractals each
with its own fractal dimension $f(a)$. So $f(a)$ describes how
densely the subsystems with the singularity exponent $a$ are
distributed. It should be noted that power law behaviors
(\ref{p1a}) and (\ref{N1a}) are the fundamental assumptions of the
multifractal analysis.

\vspace{3mm}

The convenient way how to keep track with $p_i$'s is to examine
the scaling of the corresponding moments. For this purpose one can
define a ``partition function" as
\begin{equation}
Z(q) = \sum_i p_i^q = \int da \ n(a) l^{-f(a)} l^{qa}\, ,
\label{part1}
\end{equation}
\noi ($n(a)$ is (weakly $l$ dependent) proportionality function
having its origin in relations (\ref{p1a}) and (\ref{N1a})).  In
the small $l$ case the asymptotic behavior of the  partition
function can be evaluated by the method of steepest descents. As a
result we get the scaling
\begin{equation}
Z(q)\sim l^{\tau}\, ,
\end{equation}
\noi with
\begin{eqnarray}
&&\tau(q) \equiv \min_{a} (q a - f(a)) =
q a_0(q) - f(a_0(q))\, , \nonumber \\
&& \Rightarrow \, f'(a_0(q)) = q \;\;\; \mbox{and} \;\;\; a_0(q) =
\tau'(q)\, . \label{LT1}
\end{eqnarray}
\noi These are precisely the Legendre transform relations.
Scaling function $\tau(q)$ is called correlation exponent or mass
exponent of the $q$th order. So for the purpose of multifractal
description we may use either of the conjugated couples $f(a_0),
a_0$ or $\tau(q), q$. For the future reference we will need to
know that $\tau(0) = -D$ and $\tau(1)= 0$ (see e.g.,\cite{Man1}).
Let us finally stress that if not stated otherwise, we will often
``abuse'' notation and write simply $a$ instead of $a_0$.

\subsection{Generalization of Eqs.(\ref{ren3}) and
(\ref{ren4}) to fractal sample spaces and multifractals}

With the definitions of (multi)fractal dimensions at hand we may
now generalize Eqs.(\ref{ren3}) and (\ref{ren4}). Let us assume
first that we have a fractal support $M$ on which is defined a
continuous PDF ${\mathcal{F}}(x)$. Following the renormalization
prescription of Section III we know that in order to obtain the
renormalized Renyi's entropy we have to know
${\cal{I}}_{\alpha}({\cal{E}}_n)$. This can be done by realizing
that the uniform distribution is now ${\cal{E}}_n = \left\{
\frac{1}{N_l}, \ldots,\frac{1}{N_l} \right\}$. Here $N_l$ is the
minimal covering (with cubes of size $l^d$) of the fractal set in
question and $n=1/l$. Due to scaling law (\ref{fra1}) the
(pre)fractal  volume $V_l = N_l l^D$ converges to the actual
(finite) fractal volume $V$ in the $l \rightarrow 0$ limit. As a
result ${\cal{E}}_n = \left\{ \frac{l^D}{V_l}, \ldots,
\frac{l^D}{V_l} \right\}$, and hence
\begin{equation}
{\cal{I}}_{\alpha}({\cal{E}}_n) = \log_2 V_l - D\log_2 l\, .
\end{equation}
\noi In the $n \rightarrow \infty$ (i.e., $l \rightarrow 0$) limit
we prove in Appendix D that either
\begin{eqnarray}
\tilde{{\cal{I}}}_{\alpha}({\cal{F}}) &\equiv& \lim_{n \rightarrow
\infty} ({\cal{I}}_{\alpha}({\cal{P}}_n)
- {\cal{I}}_{\alpha}({\cal{E}}_n) ) \nonumber \\
& = & \frac{1}{(1-\alpha)} \, \log_2 \left(\frac{\int_M d \mu \,
{\cal{F}}^{\alpha}({\bf{x}})}{\int_M d \mu \, 1/V^{\alpha}}
\right) \, , \label{ren55}
\end{eqnarray}
\noi or
\begin{eqnarray}
{\cal{I}}_{\alpha}({\cal{F}}) &\equiv& \lim_{n \rightarrow \infty}
({\cal{I}}_{\alpha}({\cal{P}}_n)
-{\cal{I}}_{\alpha}({\cal{E}}_n)|_{V=1}
)\nonumber \\
&=& \lim_{n \rightarrow \infty} ({\cal{I}}_{\alpha}({\cal{P}}_n) +
D \log_2 l) \nonumber \\ &=& \frac{1}{(1-\alpha)} \log_2 \left(
\int_M d \mu \, {\cal{F}}^{\alpha}({\bf{x}}) \right)\, ,
\label{ren6}
\end{eqnarray}
\noi in conformity with the chosen renormalization prescription.
The measure $\mu $ is the Hausdorff measure. Note that the RHS's
of (\ref{ren55}) and (\ref{ren6}) are finite provided the integral
$\int_M d  \mu \,{\cal{F}}^{\alpha}({\bf{x}})$ exists.
From (\ref{ren6}) the asymptotic expansion (\ref{D01}) for
${\cal{I}}_{\alpha}({\cal{P}}_n)$ reads
\begin{eqnarray}
{\cal{I}}_{\alpha}({\cal{P}}_n) = D\log_2 n +
{\mathcal{I}}_{\alpha}({\mathcal{F}}) + o(1)\, .
\end{eqnarray}
This means that $d(\alpha)$ defined in (\ref{D1}) boils down to
\begin{equation}
d(\alpha) = \lim_{n\rightarrow\infty}
\frac{{\cal{I}}_{\alpha}({\cal{P}}_n)}{\log_2 n} = D\, , \;\;\;
\mbox{for} \; \forall \, \alpha\,. \label{eq.4.12}
\end{equation}
We remark that the information measure $D \log_2 n$ appearing in
(\ref{ren45}) and (\ref{ren6}) is nothing but an
information--theoretical analogue of the Boltzmann entropy: $S =
k_B \ln W$ ($k_B$ is the Boltzmann constant and $W$ is the number
of accessible microstates). This is so because both
${\mathcal{I}}_{\alpha}({\mathcal{E}}_n)$ ($=
{\mathcal{H}}({\mathcal{E}}_n)$ for $\forall$ $\alpha$) and the
Boltzmann entropy $S$ describe systems where all possible outcomes
(or accessible microstates) have assigned equal probabilities
(constant PDF). Thus ${\mathcal{I}}_{\alpha}({\mathcal{E}}_n)$
alike $S$ are both maximal attainable entropies compatible with a
given set of all possible outcomes (or accessible microstates).

\vspace{3mm}

%
%
Foregoing analysis can be also utilized to multifractals. In fact,
by employing the multifractal measure~\cite{Fed1}
\begin{eqnarray}
\mu_{{\mathcal{P}}}^{(\alpha)}(d;l) =
\sum_{k\mbox{\footnotesize{th box}}} \frac{p^{\alpha}_{nk}}{l^d} \
\stackrel{l \rightarrow 0}{\longrightarrow} \ \left\{
\begin{array}{ll}
0 & \mbox{if $d < \tau(\alpha)$}\\
\infty & \mbox{if $d > \tau(\alpha)$}\, ,
\end{array}
\right. \label{mes2}
\end{eqnarray}
we prove in Appendix F that
\begin{eqnarray}
{\mathcal{I}}_{\alpha}(\mu_{{\mathcal{P}}}) &\equiv& \lim_{l
\rightarrow 0} \ \left({\mathcal{I}}_{\alpha}({\mathcal{P}}_n) -
{\mathcal{I}}_{\alpha}({\mathcal{E}}_n) |_{V=1} \right)\nonumber \\
&=& \lim_{l \rightarrow 0} \
\left({\mathcal{I}}_{\alpha}({\mathcal{P}}_n) +
\frac{\tau(\alpha)}{(\alpha -1)} \log_2 l \right) \nonumber \\
&=& \frac{1}{(1-\alpha)} \ \log_2\left( \int_a
d\mu_{{\mathcal{P}}}^{(\alpha)}(a)\right)\, . \label{mes6}
\end{eqnarray}
\noi Eq.(\ref{mes6}) implies the asymptotic expansion
\begin{equation}
{\mathcal{I}}_{\alpha}({\mathcal{P}}_n) =
\frac{\tau(\alpha)}{(\alpha -1)} \log_2 n +
{\mathcal{I}}_{\alpha}(\mu_{{\mathcal{P}}}) + o(1)\, .
\end{equation}
\noi Consequently we note that $d(\alpha)$  of (\ref{D1}) reads
\begin{equation}
d(\alpha) = \lim_{n \rightarrow \infty}
\frac{{\mathcal{I}}_{\alpha}({\mathcal{P}}_n)}{\log_2 n} =
\frac{\tau(\alpha)}{(\alpha -1)}\, . \label{b7}
\end{equation}
Unlike in fractal sample spaces, in multifractals $d(\alpha)$
depends on $\alpha$. Note that in the case of smooth PDF's the
integrated probability $p_i(l)$ scales as $l^{f(a)}$ and so we
have a unifractal characterized by a single dimension $a =
f(a)\equiv D$. This implies that $\tau/(\alpha-1) = D$ and hence
for smooth PDF's we naturally recover the result (\ref{eq.4.12}).
It should be emphasized that when the outcome space is a simple
metric space (like ${\mathbb{R}}^d$) then it is known that the
fractal dimension $D$ coincides with the usual topological
dimension\cite{Man1,Fed1} and so, for instance, $D=d$ in the case
of ${\mathbb{R}}^d$.

\subsection{Generalized dimensions and reconstruction
theorem}\label{rec}

After this brief intermezzo we now turn back to the question
whether there is any connection of R\'{e}nyi's entropy with
(multi)fractal systems.  At present it seems to us that there are
at least two such connections. The first, more formal connection,
is associated with the so called  generalized dimensions of the
$q$th order defined as:
\begin{eqnarray}
{\cal{D}}_q \equiv \lim_{l\rightarrow 0} \left( \frac{1}{(q-1)}
\frac{\ln Z_q}{\ln l} \right) = \frac{\tau(q)}{(q-1)}\, .
\label{r-gd}
\end{eqnarray}
\noi In passing the reader should notice that ${\cal{D}}_q$ is
nothing but $d(\alpha = q)$ introduced in (\ref{b7}).  A complete
knowledge of the collection of generalized dimensions
${\cal{D}}_q$ is equivalent to a complete physical
characterization of the fractal\cite{HaPr1}. It should be noted in
this connection that the fractal dimension, the information
dimension and the correlation dimension (all frequently used in
the deterministic chaotic systems\cite{Ruel1}) are, respectively
${\cal{D}}_0$, ${\cal{D}}_1$ and ${\cal{D}}_1$. In fact, all
${\cal{D}}_q$ are necessary to describe uniquely general fractals
e.g., strange attractors\cite{HaPr1}. This is analogous to
statistical physics where one needs all cumulants to get the full
density matrix. Mathematically this corresponds to Hausdorff's
moment problem\cite{Wid1}.

\vspace{3mm}

While the proof in\cite{HaPr1} is based on a rather complicated
self--similarity argumentation we can understand the core of this
assertion using a different angle of view. In fact, employing the
information theory we will show that the assumption of a
self--similarity is not really fundamental and that the conclusion
of\cite{HaPr1} has more general applicability. For this purpose
let us define the information--distribution function of
${\cal{P}}$ (see e.g., \cite{Re2}) as
\begin{equation}
{\cal{F}}_{\cal{P}}(x) = \sum_{-\log_2 p_k < x} p_k \, .
\label{Fp1}
\end{equation}
\noi The latter represents the total probability carried out by events
with information contents ${\cal{I}}_k = -\log_2 p_k < x$.
Note also that for $x<0$ the sum in (\ref{Fp1}) is empty and so
${\cal{F}}_{\cal{P}}(x)=0$. Realizing that
\begin{eqnarray*}
2^{(1-\alpha)x}d{\cal{F}}_{\cal{P}}(x) \ \approx \!\!\!\!
\sum_{x\leq {\cal{I}}_k < x+dx} 2^{(1-\alpha){\cal{I}}_k} p_k \ =
\!\!\!\! \sum_{x\leq {\cal{I}}_k < x+dx} p_k^{\alpha} \, ,
\end{eqnarray*}
\noi we may write
\begin{equation}
{\cal{I}}_{\alpha}({\cal{P}}) = \frac{1}{(1-\alpha)} \log_2 \left(
\int_{x=0}^{\infty} \ 2^{(1-\alpha)x}d{\cal{F}}_{\cal{P}}(x)
\right)\, .
\label{Lap1}
\end{equation}
\noi The former integral should be understood in the Stieltjes
sense (${\cal{F}}_{\cal{P}}(x)$ is generally discontinuous).
Taking the inverse Laplace--Stiltjes transform of (\ref{Lap1}) we
obtain
\begin{eqnarray}
{\cal{F}}_{\cal{P}}(x) &=& \frac{1}{2\pi i} \int_{-i \infty +
\sigma}^{i \infty + \sigma} dp \;
\frac{e^{px} \ e^{-p{\cal{I}}_{\alpha}({\cal{P}})}}{p}\nonumber \\
 &=& \sum_l  \frac{p_l}{2\pi i} \int_{-i \infty + 0_+}^{i \infty
+ 0_+} dp \, \frac{e^{p(x + \log_2 p_l)}}{p} \, ,
\label{LST1}
\end{eqnarray}
\noi with $p\! = \! (\alpha-1)\ln2 $. The constant $\sigma $ is
dictated by requirements that it should be positive and that all
singularities of $\frac{e^{-p{\cal{I}}_{\alpha}}}{p}$ should lie
to the left of the vertical line $\Re(p) = \sigma$ in the complex
$p$--plane. As $e^{-p{\cal{I}}_{\alpha}}$ is basically $\sum_k
p_k^{\alpha}$ it means that $\frac{e^{-p{\cal{I}}_{\alpha}}}{p}$
is analytic on the half--plane $\{ p ~| \Re(p) > 0 \}$. As a
result we may choose $\sigma = 0_+$. For $(x+\log_2 p_k) < 0$ we
may close the contour by a semicircle in the right half of the
plane. In this region integrand is analytic and so
${\cal{F}}_{\cal{P}}(x) = 0$ as it should be. For $(x + \log_2
p_k)>0$, the semicircle must be placed in  the left half plane,
which yields then correct ${\cal{F}}_{\cal{P}}(x)$ of
Eq.(\ref{Fp1}).

\vspace{3mm}

Disadvantage of the inverse formula (\ref{LST1}) is that $p$ (and
so $\alpha$) gets its values from ${\mathbb{C}}$, or more
specifically, one needs (at best) all complex $p$'s belonging to
the small circle around $p =0$ to reconstruct the underlying
distribution. It is however clear that in order to determine how
many $\alpha$'s are really needed to fully reconstruct
${\mathcal{P}}$ one must resort to the real inverse Laplace
transform instead. Such a reversal indeed exists and is provided
by, the so called, Widder--Stieltjes inverse formula\cite{Wid1}:
\begin{displaymath}
{\cal{F}}_{\cal{P}}(x) \approx \sum_{n=0}^{\Lambda} \frac{\left(
-\frac{\Lambda}{x}\right)^n}{n!} \, \left[ \exp\left(-
\frac{\Lambda}{x} \ {\cal{I}}_{\left(\Lambda/\ln(2)x +1
\right)}\right)\right]^{(n)}\, ,
\end{displaymath}
\noi or (after setting $\frac{\Lambda}{x} = z$)
\begin{equation}
{\cal{F}}_{\cal{P}}\left(\frac{\Lambda}{z}\right) \approx
\sum_{n=0}^{\Lambda} \frac{\left( -z \right)^n}{n!} \, \left[
\exp\left(- z \ {\cal{I}}_{\left(z/\ln(2) +1
\right)}\right)\right]^{(n)}\, , \label{eq.4.28}
\end{equation}
\noi here $\Lambda$ is a regulator which has to be set to
$+\infty$ at the end of calculations. It is important to recognize
that the RHS of (\ref{eq.4.28}) depends on all $\alpha \in
[1,\infty)$. Other, more intuitive, proof of the same fact is
provided in Appendix G. In addition, in Appendix H we show that a
similar ``reconstruction'' theorem holds also for THS entropy
${\mathcal{S}}_{\alpha}$.

\vspace{3mm}

As a result, when working with ${\cal{I}}_{\alpha}$ of different
orders we receive more information than restricting our
consideration to only one $\alpha$. In this connection it is
illuminating to rewrite the complex integral in (\ref{LST1}) as
\begin{eqnarray}
&&\int_{-i\infty + 0_+}^{i\infty + 0_+} dp \ \frac{e^{p(x+ \log_2
p_k)}}{p}\nonumber \\ &&\mbox{\hspace{1.5cm}}= \mbox{PP}
\int_{-\infty}^{\infty} dp \ \frac{e^{ip(x + \log_2 p_k)}}{p} +
i\pi\, . \label{b8}
\end{eqnarray}
\noi Here PP stands for the principal part (associated to the pole
at $p=0$). The term $i\pi$ is the sole contribution from $p=0$
(i.e., $\alpha =1$), while $\mbox{PP}(\ldots)$ part corresponds to
the contribution from the (imaginary axis) neighborhood of $p=0$.
In the case when $(x + \log_2 p_k) > 0$ then $\mbox{PP}(\ldots) =
i\pi$ and when $(x+ \log_2 p_k)<0$ then $\mbox{PP}(\ldots) =
-i\pi$, so the $\alpha =1$ contribution has precisely $50\%$
dominance. It should be also realized that $\mbox{PP}(\ldots)$ is
ruled for most $p_k$'s by $p$'s from the close proximity of $p=0$.
In fact,
\begin{eqnarray}
&&\mbox{PP} \int_{-\infty}^{\infty} dp \ \frac{e^{ip(x + \log_2
p_k)}}{p}\nonumber \\
&&\mbox{\hspace{10mm}} = \ \mbox{PP} \int_{-\delta}^{\delta} dp \
\frac{e^{ip(x +
\log_2 p_k)}}{p} - 2i \ \mbox{si}(\delta y)\nonumber \\
&&\mbox{\hspace{10mm}}\approx \ \mbox{PP} \int_{-\delta}^{\delta}
dp \ \frac{e^{ip(x + \log_2 p_k)}}{p}\nonumber \\
&& \mbox{\hspace{15mm}} + \ 2i \varepsilon (y)\left( \pi/2 -
\delta |y| + {\mathcal{O}}((\delta |y|)^3)\right)\, ,
\end{eqnarray}
\noi with $\delta$ being the $\delta$--neighborhood of $p=0$,
si$(x)$ being the sine integral and $y = (x + \log_2 p_k)$. Hence
we see that when the outcome space is a discrete set we need
generally all ${\mathcal{I}}_{\alpha}$'s with $\alpha \in [1,
\infty)$ to determine ${\mathcal{P}}$ albeit the most dominant
contribution comes from the relatively small neighborhood of
${\mathcal{I}}_1 = {\mathcal{H}}$. The latter statement is the
discrete--space
variant of the conclusion
in\cite{HaPr1}.

\vspace{3mm}

Let us now briefly comment on the reconstruction theorem for the
cases when the outcome space is a $d$--dimensional subset of
${\mathbb{R}}^d$. By covering the subset with the mesh of
$d$--dimensional cubes of size $l^d = 1/n^d$ we obtain similarly
as in Section~III the integrated distributions ${\mathcal{P}}_n =
\{p_{nk} \}$ and ${\mathcal{E}}_n = \{ {\mathcal{E}}_{nk} \}$. The
corresponding information--distribution function now reads
\begin{eqnarray}
{\mathcal{F}}_{{\mathcal{P}}_n/{\mathcal{E}}_n}(x) &=&
\sum_{-\log_2(p_{nk}/{\mathcal{E}}_{nk})< \ x} \!\!\!\!\!\!\!\!\!
\left(p_{nk}/{\mathcal{E}}_{nk}\right)/\sum_k
\left(p_{nk}/{\mathcal{E}}_{nk}\right)\nonumber \\
&=&  \sum_{-\log_2(p_{nk}/{\mathcal{E}}_{nk})< \ x}
\!\!\!\!\!\!\!\!\! \left(p_{nk}/{\mathcal{E}}_{nk}\right)\
\frac{1}{V n^d} \, .
\end{eqnarray}
\noi This implies (for $V=1$) that
\begin{displaymath}
\int_{x= - d \log_2 n}^{\infty} \!\!\!\!\!\!\!\!\!\!\!\! \!\!\!\!
2^{(1-\alpha)x}\
d{\mathcal{F}}_{{\mathcal{P}}_n/{\mathcal{E}}_n}(x) = \frac{\sum_k
p_{nk}^{\alpha}}{\sum_k {\mathcal{E}}_{nk}^{\alpha}}\, ,
\end{displaymath}
\noi and so in accord with (\ref{ren2})
\begin{eqnarray}
&&{\mathcal{I}}^{(n)}_{\alpha}({\mathcal{F}}) =
\frac{1}{(1-\alpha)}\ \log_2 \left(\int_{x = -d \log_2
n}^{\infty}\!\!\!\!\!\!\!\!\!\!\!\!\!\!\!\! 2^{(1-\alpha) x} \ d
{\mathcal{F}}_{{\mathcal{P}}_n/{\mathcal{E}}_n}(x)
\right)\, ,\nonumber \\
&&{\mathcal{I}}_{\alpha}({\mathcal{F}}) = \lim_{n \rightarrow
\infty}{\mathcal{I}}^{(n)}_{\alpha}({\mathcal{F}}) \, .
\end{eqnarray}
\noi Using the Widder--Stiltjes inverse formula we may re--create
${\mathcal{F}}_{{\mathcal{P}}_n/{\mathcal{E}}_n}(x)$ (and hence
${\mathcal{F}}$) in terms of
${\mathcal{I}}^{(n)}_{\alpha}({\mathcal{F}})$'s. But the important
moral here is that in the continuous limit (large $n$) $x\in
(-\infty, \infty)$ and so $\alpha \in (-\infty, \infty)$. Unlike
in discrete sample spaces, all ${\mathcal{I}}_{\alpha}$, including
those with $\alpha < 1$, are needed now to pinpoint the underlying
PDF.

\vspace{3mm}

It should be born in mind that from a purely mathematical point of
view the reconstruction procedure presented here is by no means
the proof which extends easily to (multi)fractal systems - there
is now obvious analogue of the Widder--Stiltjes inverse formula
there. It should be rather taken as an indication that in general
systems all ${\cal{I}}_{\alpha}$ with $\alpha \in (-\infty,
\infty)$ are needed to determine uniquely the probability
distribution. This is basically a weak version of the celebrated
{\em moment problem of Hausdorff}\cite{Wid1}. The latter resonates
with the finding that for deterministic chaotic systems the
multifractal scaling function $\tau(q)$ often exists even for
negative values of $q$. In those cases the partition function
(\ref{part1}) is dominated by very small values of $p_i$. Hence
one may be skeptical about the real existence of such a
negative--$q$ scaling behavior since the latter can be easily
disrupted by fluctuations. In fact, if we explore the stability of
Renyi's entropy for negative $\alpha$ by adding a small imaginary
part into $\alpha$ we obtain Fig.\ref{fig5}.
\begin{figure}[h]
\vspace{4mm} \epsfxsize=8cm \centerline{\epsffile{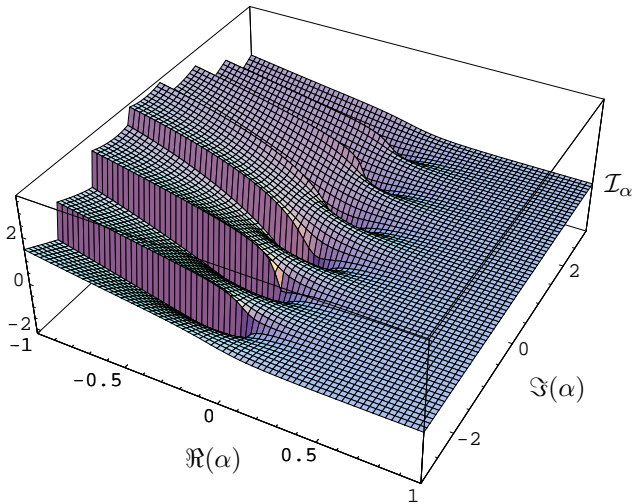}}
\vspace{4mm} \caption{\em  A plot of Renyi's entropy
$\mathcal{I}_{\alpha}({\mathcal{P}})$ for 2 dimensional
${\mathcal{P}} = (p_1,p_2) = (p, 1-p)$. We choose $p =0.01$.}
\label{fig5}
\begin{picture}(20,7)
\put(205,85){ $\Im(\alpha)$ } \put(75,58){ $\Re(\alpha)$ }
\put(237,162){$\mathcal{I}_{\alpha}$}
\end{picture}
\end{figure}

\noi As $p$ goes closer to zero there is a violent proliferation
of branch cuts in the left half of the complex $\alpha$--plane. So
information conveyed by Renyi's entropy with negative $\alpha$
starts to be highly unreliable. Because R{\'e}nyi's entropy is
connected with the generalized dimensions via relation
(\ref{r-gd}) such a breakdown of scaling for negative $q$'s (and
hence $\alpha$'s) should be inevitable in various deterministic
chaotic systems. This is indeed the case, see e.g.,\cite{RCBa}.

\vspace{3mm}

The former reasonings may, to a certain extent, vindicate the use
of $\alpha \geq 0$ in usual information theory. The bound $\alpha
\geq 0$ can be hence merely understood as a reliability bound
imposed on the conveyed information.


\subsection{Thermodynamic formalism and MaxEnt}

The second connection which we intend to advocate and progress
here is the connection with the maximal entropy principle
(MaxEnt). We will show that from the MaxEnt point of view,
extremizing Shannon's entropy on (multi)fractals is equivalent to
extremizing directly R\'{e}nyi's entropy without invoking the
(multi)fractal structure explicitly.
An  explicit illustration of this point on
the network of cosmic strings will be given elsewhere.

\vspace{3mm}

Consider a support paved with boxes of size $l$ and let the
integrated probability in the $k$th box is denoted as $p_k$.
Shannon's entropy of such a process is then
\begin{displaymath}
{\mathcal{I}} = - \sum_k p_k(l) \log_2 p_k(l)
\end{displaymath}
\noi The important observation of the multifractal theory is that
for $q=1$
\begin{equation}
a(1) = \frac{d \tau(1)}{dq} = \lim_{l \rightarrow 0} \frac{\sum_k
p_k(l) \log_2 p_k(l)}{\log_2 l}\, .
\end{equation}
\noi It can be shown that the number $a(1) = f(a(1))$ describes
the Hausdorff--Besicovich dimension of the set on which the
probability is concentrated (see e.g,\cite{Fed1}). This means that
the probability distribution ${\mathcal{P}}_n$ is cumulated on the
$l$--mesh cubes with $p_k(l) \sim l^{a(1)}$. In fact, the relative
probability of the complement set approaches zero in the
$l\rightarrow 0$ limit\cite{Fed1}. This statement goes also under
the name {\em Billingsley theorem}
\cite{Bi1} or {\em curdling}\cite{Man1}. The corresponding subset
${\mathcal{M}}$ is known as the {\em measure theoretic support}.
Let us thus write
\begin{eqnarray}
d_H ({\mathcal{M}}) \equiv f(a(1) ) &=& \lim_{l \rightarrow 0}
\frac{1}{\log_2 l} \, \sum_k p_k(l) \log_2 p_k(l) \nonumber \\
&\approx& \, \frac{1}{\log_2 \varepsilon} \, \sum_k p_k(\varepsilon)
\log_2 p_k(\varepsilon)\, .
\end{eqnarray}
\noi Here $\varepsilon$ corresponds to a cutoff (or coarse
graining) scale of the grid. For the further convenience we will
keep $\varepsilon = l_{cut}$ finite throughout all our
calculations and set $\varepsilon \rightarrow 0$ only at the end.

\vspace{3mm}

In the case of multifractal systems one is often interested in
entropy of only certain (uni)fractal subsets. For such a purpose
it is useful to introduce a one--parametric family of normalized
distributions (zooming or escort distributions) $\varrho(q)$ as
\begin{displaymath}
\varrho_i(q,l) = \frac{[p_i(l)]^q}{\sum_j [p_j(l)]^q} \, \sim \,
l^{qa_i - \tau} = l^{f(a_i)}\, .
\end{displaymath}
\noi Because the distribution $\varrho(q,l)$ alters the scaling of
the original distribution ${\mathcal{P}}_n$, the corresponding
measure theoretic support will change. As a mater of fact,
distribution $\varrho(q,l)$ enables to form an ensemble of measure
theoretic supports ${\mathcal{M}}^{(q)}$ parametrized by $q$.
Parameter $q$ provides a ``zoom in"  mechanism to probe various
regions of a different singularity exponent. Indeed, from
(\ref{LT1}) we have
\begin{eqnarray}
&& df(a) = \left\{ \begin{array}{ll}
                          \leq da & \mbox{if $q\leq1$}\\
                          \geq da & \mbox{if $q\geq1$}\, .
                          \end{array}\right.
\label{zoom1}
\end{eqnarray}
\noi Integrating (\ref{zoom1}) from $a(q=1)$ to $a$ we obtain
\begin{eqnarray}
&& f(a) = \left\{ \begin{array}{ll}
                          \leq a & \mbox{if $q\leq1$}\\
                          \geq a & \mbox{if $q\geq1$}\, ,
                          \end{array}\right.
\end{eqnarray}
%
\noi and so
 for $q >1$ $\varrho(q)$ puts emphasis on the more
singular regions of ${\mathcal{P}}_n$, while for $q <1$ the
accentuation is on the less singular regions (see also Fig.2).
\begin{figure}[h]
\vspace{4mm} \epsfxsize=8cm \centerline{\epsffile{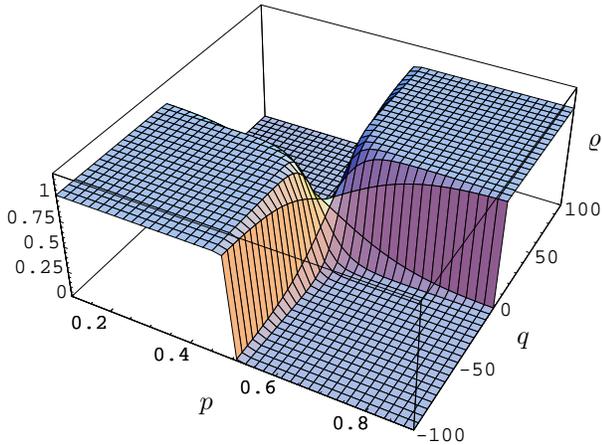}}
\vspace{4mm} \caption{\em  A plot of the zooming distribution for
2 dimensional ${\mathcal{P}}$: $\varrho(q) = p^q/(p^q +
(1-p)^q)$.} \label{fig3}
\begin{picture}(20,7)
\put(200,80){ $q$ } \put(80,55){ $p$ } \put(230,155){$\varrho$}
\end{picture}
\end{figure}
%
The corresponding
%
%
fractal dimension of the measure theoretic support
${\mathcal{M}}^{(q)}$ of $\varrho(q)$ is
\begin{eqnarray}
d_{H}({\mathcal{M}}^{(q)}) &=& \lim_{l\rightarrow 0}
\frac{1}{\log_2 l} \, \sum_k \varrho_k(q,l) \log_2
\varrho_k(q,l)\nonumber \\
&\approx& \, \frac{1}{\log_2 \varepsilon} \, \sum_k
\varrho_k(q,\varepsilon) \log_2 \varrho_k(q,\varepsilon)\, .
\label{mts1}
\end{eqnarray}
\noi We can now use (\ref{mts1}) to find the promised connection
between multifractals and R\'{e}nyi's entropy. To do this let us
observe that the curdling (\ref{mts1}) mimics
 the situation
occurring
 in equilibrium statistical physics. There in {\em
canonical formalism} one works with (usually infinite) ensemble of
identical macroscopic systems with all possible energy
configurations. Notwistanding only the configurations with $E_i =
\langle E \rangle$ dominate in thermodynamic limit. In fact,
defining the ``microcanonical" partition function
%
%
%
\begin{displaymath}
Z_{mic} = \left(\sum_{a_k \in (a_i, a_i + d a_i)} \!\!\!\! 1
\right) = dN(a_i)\, ,
\end{displaymath}
\noi one gets for $a_i \approx \log_2 (p_i)/\log_2 \varepsilon$
(c.f., (\ref{p1a}))
\begin{eqnarray}
&&\langle a \rangle_{mic} \ = \ \!\!\!\!\!\!\sum_{a_k \in (a_i,
a_i + da_i)} \frac{a_k}{Z_{mic}}
\ \approx \ a_i\, , \nonumber \\
&&\langle f(a) \rangle_{mic} \ = \ \!\!\!\!\!\!\sum_{a_k \in (a_i,
a_i + da_i)} \frac{f(a_k)}{Z_{mic}} \ \approx \ f(a_i)\, .
\end{eqnarray}
\noi Because in the micro--canonical approach the distribution is
uniform (${\mathcal{E}}(a_i)= \{1/dN(a_i)\}$), the corresponding
Shannon--Gibbs entropy boils down to the micro--canonical (or
Boltzmann) entropy
\begin{equation}
{\mathcal{H}}({\mathcal{E}}(a_i)) = \log_2 dN(a_i) = \log_2
Z_{mic}\, ,
\end{equation}
\noi and hence
\begin{equation}
\frac{{\mathcal{H}}({\mathcal{E}}(a_i))}{\log_2 \varepsilon}\
\approx \ - \langle f(a) \rangle_{mic}\, .
\end{equation}
\noi Interpreting  $E_i = -a_i \log_2 \varepsilon$ as ``energy''we
may define the ``inverse temperature" $1/T = \beta /\ln2$ (note
that $k_B = 1/\ln2 $ here) as
\begin{eqnarray*}
1/T= \left.\frac{\partial {\mathcal{H}}}{\partial
E}\right|_{E=E_i} = -\frac{1}{\ln \varepsilon \ Z_{mic}}\
\frac{\partial Z_{mic}}{\partial a_i} = f'(a_i) =  q \, .
\end{eqnarray*}
\noi Legendre transform then allows to determine the conjugate
function $\tau(q)$ via
\begin{equation}
\langle f(a) \rangle_{mic} \approx q \langle a \rangle_{mic}
-\tau(q)\, .
\end{equation}
On the other hand, defining the ``canonical" partition function as
\begin{displaymath}
Z_{can} = \sum_i p_{i}(\varepsilon)^{q} = \sum_i e^{-\beta E_i}\,
,
\end{displaymath}
\noi (where the identifications $\beta=q \ln2$ and
$E_i=-\log_2(p_i(\varepsilon)) $ are made) the corresponding means
are
\begin{eqnarray}
&& a(q) \equiv \langle a \rangle_{can} = \sum_{i}
\frac{a_i}{Z_{can}} e^{-\beta E_i}\nonumber \\
&&\mbox{\hspace{2cm}}\approx \frac{\sum_i \varrho_i
(q,\varepsilon)\log_2
p_i(\varepsilon)}{\log_2 \varepsilon} \, , \nonumber \\
&& f(q) \equiv \langle f(a) \rangle_{can} = \sum_i
\frac{f(a_i)}{Z_{can}} e^{-\beta E_i}\nonumber \\
&&\mbox{\hspace{2.5cm}} \approx \frac{\sum_i
\varrho_i(q,\varepsilon) \log_2 \varrho_i(q,\varepsilon)}{\log_2
\varepsilon}\, . \label{can2}
\end{eqnarray}
\noi Let us observe two things. Firstly, the fractal dimension of
the measure theoretic support $d_H({\mathcal{M}}^{(q)})$ is simply
$f(q)$. If $q$ is a solution of the equation $a_i = \tau'(q)$ then
in the ``thermodynamic" limit ($\varepsilon \rightarrow 0$) we can
identify
\begin{eqnarray}
&& a(q) = \langle a \rangle_{can} = \langle a \rangle_{mic}
\approx a_i\, ,\nonumber \\
&& f(q) = \langle f(a) \rangle_{can} = \langle f(a) \rangle_{mic}
\approx f(a_i)\, . \label{thlim1}
\end{eqnarray}
\noi Eqs.(\ref{can2}) then provide a parametric relationship
between $f(q)$ and the singularity exponent $a(q)$. When the
parameter $q$ is eliminated one recovers the usual singularity
spectrum $f(a)$. Eqs.(\ref{can2}) imply that $\langle f
\rangle_{can} = q\langle a \rangle_{can} - \tau$, $\langle a
\rangle_{can}= d\tau /dq$, and so again the Legendre transform
applies. Secondly, because the micro--canonical and canonical
entropies coincide in the thermodynamic limit
%
\begin{equation}
{\mathcal{H}}({\mathcal{E}}(a)) \approx - \sum_k
\varrho_k(q,\varepsilon) \log_2 \varrho_k(q,\varepsilon) \nonumber
\equiv {\mathcal{H}}({\mathcal{P}}_n)\mbox{{$|$}}_{f(q)}\, .
\end{equation}
\noi Here we have used the subscript $f(q)$ to emphasize that the
Shannon entropy ${\mathcal{H}}({\mathcal{P}}_n)$ is basically the
entropy of an unifractal specified by the fractal dimension $f(q)$
defined in (\ref{can2}).
%
%
Because of relations (\ref{thlim1}) and the Legendre transform
(\ref{LT1}) we obtain after a short algebra
\begin{eqnarray}
\frac{\left.{\mathcal{H}}({\mathcal{P}}_n)\right|_{f(q)} }{\log_2
\varepsilon } + f &=& \frac{{\mathcal{I}}_q}{\log_2 \varepsilon} +
\frac{\tau}{q-1}\nonumber \\
&~&\nonumber \\
 & -& q\left[ \left(\tilde{a} - \langle a \rangle_{can}\right) + \frac{(\tilde{\tau} -
 \tau)}{1-q} \right]
\, , \label{sh24}
\end{eqnarray}
\noi with $q$ determined by the condition $\tau'(q) = a$ and
\begin{eqnarray*}
&&\tilde{a} =  \frac{\sum_i \varrho_i (q,\varepsilon)\log_2
p_i(\varepsilon)}{\log_2 \varepsilon}\, , \;\;\;\;\; \tilde{\tau}
= \frac{\log_2 \sum_i p_i^q(\varepsilon)}{\log_2 \varepsilon}\, .
\end{eqnarray*}
Applying l'Hospital's rule we find that
\begin{equation}
\lim_{\varepsilon\rightarrow 0}\left[ \left(\tilde{a} - \langle a
\rangle_{can}\right) + \frac{(\tilde{\tau} -
 \tau)}{1-q} \right]\log_2 \varepsilon =0\, .
\end{equation}
%
%
%
%
%
\noi Multiplying (\ref{sh24}) by $\log_2 \varepsilon$, taking the
small $\varepsilon$ limit and employing the renormalization
prescriptions (\ref{ren6}) and (\ref{mes6}) we finally receive
that
\begin{equation}
{\mathcal{I}}_q^r = {\mathcal{H}}^r\mbox{\large{$|$}}_{f(q)}\, .
\label{shren1}
\end{equation}
\noi The superscript $r$ indicates the renormalized quantities. To
understand  (\ref{shren1}) let us note that
${\mathcal{H}}({\mathcal{P}}_n)\mbox{\large{$|$}}_{f(q)}$ can be
alternatively written as
\begin{eqnarray}
{\mathcal{H}}({\mathcal{P}}_n)\mbox{\large{$|$}}_{f(q)} &\approx&
\sum_{k=1}^{dN(a)} \frac{p_k(\varepsilon)}{ \sum_{l=1}^{dN(a)}
p_l(\varepsilon)}\ \log_2 \left(  \frac{p_k(\varepsilon)}{
\sum_{l=1}^{dN(a)} p_l(\varepsilon)} \right)\nonumber \\
&~& \nonumber \\ &=& \log_2 dN(a)\, .\label{shren2}
\end{eqnarray}
\noi Denoting the incomplete distribution $\sum_{k=1}^{dN(a)}
p_k(\varepsilon)$ $< 1$ as ${\mathcal{S}}$ and the conditional
distribution $ \left\{ p_k(\varepsilon)/{\mathcal{S}};\right.$
$\left. \, k = 1, \ldots, dN(a) \right\}$ as ${\mathcal{P}}'_n$
then
\begin{eqnarray}
{\mathcal{H}}({\mathcal{P}}_n)\mbox{\large{$|$}}_{f(q)}\,
&\approx& \,\,
{\mathcal{H}}({\mathcal{P}}'_n) \nonumber \\
&=& \, \frac{\sum_{k=1}^{dN(a)} p_k(\varepsilon) \log_2
p_k(\varepsilon)}{\sum_{l=1}^{dN(a)} p_l(\varepsilon)} - \log_2
\frac{1}{{\mathcal{S}}} \, . \label{shren3}
\end{eqnarray}
\noi So the RHS of (\ref{shren1}) equals to Shannon's information
of an incomplete distribution\cite{Re1,Re2} minus information
corresponding to the total probability of the incomplete system
(i.e., unifractal).
%
%

\vspace{3mm}

In passing we can observe that for $q=1$ the LHS of (\ref{shren1})
represents the Shannon entropy of the entire multifractal system,
while the RHS stands for the Shannon entropy of the unifractal
with the fractal dimension $a(1)=f(a(1))=D$. It is of course
Billingsley's theorem which makes sure that both sides match in
the continuous limit. Now, the passage from multifractals to
single--dimensional statistical systems is done by assuming that
the $a$--interval gets infinitesimally narrow and that PDF is
smooth. In such a case both $a$ and $f(a)$ collapse to $a = f(a)
\equiv D$ and $q = f'(a) =1$. So, for instance, for a statistical
system with a smooth measure and the support space
${\mathbb{R}}^d$ Eq.(\ref{shren1}) constitutes a trivial identity.
We believe that this is the primary reason why Shannon's entropy
plays such a predominant role in physics of single--dimensional
sets.

\vspace{3mm}


Let us make finally one more observation. If we apply the MaxEnt
approach to a single unifractal (say that with the dimension
$f(q)$) and try to infer the most probable incomplete distribution
which complies with whatever macroscopic constraints we know about
the unifractal subsystem, we have to look for a conditional
extremum of Shannon's entropy
${\mathcal{H}}({\mathcal{P}}_n)\mbox{\large{$|$}}_{f(q)}$. This
can be done, at least in principle, in two ways. We can either
extremize
${\mathcal{H}}({\mathcal{P}}_n)\mbox{\large{$|$}}_{f(q)}$ with the
incomplete distribution keeping ${\mathcal{S}}$ fixed, or
extremize
${\mathcal{H}}({\mathcal{P}}_n)\mbox{\large{$|$}}_{f(q)}$ directly
with respect to the zooming distribution $\varrho(q,\varepsilon)$.
The second way is often more manageable. As a result we obtain
that the least biased incomplete probability distribution on the
unifractal characterized by the dimension $f(q)$ is obtained via
extremizing R\'{e}nyi's entropy ${\mathcal{I}}_q({\mathcal{P}}_n)$
with respect to the zooming distribution $\varrho(q,\varepsilon)$.
So by changing the $q$ parameter at R\'{e}nyi's entropy one can
``skim over" all unifractal Shannon's entropies. If, additionally,
the macroscopic constraints correspond to state variables then
MaxEnt approach naturally allows for a thermodynamic description
of multifractals.
\section{Final remarks}

It was the aim of this paper to present a self--contained
discussion of R\'{e}nyi's entropy. Apart from formal information
theory aspects of R\'{e}nyi's entropy we have studied its bearing
on various topics of current interest in physics. These include
the THC non--extensive entropy, fractal and multifractal systems,
PDF reconstruction theorem, chaotic dynamical systems and MaxEnt
approach to thermodynamics.

\vspace{3mm}

It should be noted that the thermodynamical or statical concept of
entropy, though deeply rooted in physics, is rigorously defined
only for equilibrium systems or, at best, for adiabatically
evolving systems. In fact, the very existence of the entropy in
thermodynamics is attributed to Carath\'{e}odory's inaccessibility
theorem\cite{Car1} and the statistical interpretation behind the
thermodynamical entropy is then usually provided via the ergodic
hypothesis\cite{Rb1,Khin2}. When one moves away from equilibrium
there are very few clues left of how one should proceed to define
entropy. In particular, there in no general concept of ergodicity
which could come into our rescue. But just what is entropy then?
It is frequently said that entropy is a measure of disorder, and
while this needs many qualifications and clarifications it is
generally believed that this does represent something essential
about it. Insistence on the former interpretation however
naturally begs for an operational prescription.
To tackle this issue we have resorted to information theory. Here
disorder is quantified in terms of missing information and the
corresponding information entropy is a measure of our ignorance
about a system in question. We feel that the latter is a natural
and conceptually very clean extension of the equilibrium concept
of entropy.
This might be
further reinforced by the fact that the information entropy stands
a full mathematical rigor. Actually, the information theory
provides a whole hierarchy od information entropies each of which
is compatible with basic axioms of information theory and theory
of probability. Such information entropies are mutually
distinguished by their order (R\'{e}nyi's parameter). It is well
known\cite{Jay2} that the information entropy of order $1$
(Shannon's entropy) can successfully reproduce the usual
equilibrium statistical physics and hence thermodynamics on a
simple metric spaces. It was one of the aims of this paper to show
that when dealing with (multi)fractal systems one needs to use
also information entropies of orders $\alpha \neq 1$ - R\'{e}nyi
entropies. In fact, because the concept of information does not
hinge on the notion of equilibrium or non--equilibrium, one may go
even further and apply information entropies into various
non--equilibrium situations (for $\alpha =1$ case see
e.g.,\cite{PJ3} and citations therein).

\vspace{3mm}

Because of this versatile nature of R\'{e}nyi's entropy we are
rather tempted to believe that THC entropy is only derived (i.e.,
not fundamental) concept in physics. We substantiate the latter by
arguing that in certain instances - e.g., rare events systems -
THS entropy is the leading order approximation to R{\'e}nyi's
entropy. In addition, because R{\'e}nyi's entropy is a monotonous
function of THS entropy all stability conditions in thermodynamics
are identical in both cases and so from thermodynamical point of
view both entropies are indistinguishable. In those cases it is a
matter of taste and/or technical convenience which one will be
applied\cite{ArAr1}. It should be also noted that in this light an
apparent non--extensivity of THS entropy could be possibly viewed
as an artificial (local) feature of much the same origin as is a
non--periodicity of leading (i.e., local) contributions to
(globally) periodic functions.


\vspace{3mm}

It should be, however, admitted that the authors see a possible
loophole for THC entropy to play a more pivotal role - i.e., to be
an autonomous (not derived) and conceptually clean construct,
similarly as, for example, Fisher's entropy\footnote{Fisher's
entropy (or information) is an important concept in parametric
statistics as it represents a measure of the amount of information
a given statistical sample contains about the parameter which
parametrizes PDF. It is well known that there is and intimate
connection between Fisher's and Shannon's\cite{Kul} (and
Reny's\cite{Re2}) entropy, yet both concepts are completely
autonomous.} is. The loophole seem to be provided by the
\textit{quantum non--locality}.
The point is that in order to obtain some breathing space for THC
entropy some of the axioms of R\'{e}nyi's entropy must be bypassed
or at least soften. The authors feel that only plausible
possibility is to violate the axiom 3 of Section~IIA with its
additivity of independent information. In fact, we have derived
the additivity of entropies for independent experiments with the
hidden assumption that experiments are independent if (and only
if) they are uncorrelated. In quantum mechanics, however, the
relationship between independent and uncorrelated is more
delicate. At present it seems that the feasible mechanism which
questions, although in a very subtle way, the equivalence between
being independent and being uncorrelated is attributed to the
quantum non--locality and, in particular the \textit{quantum
entanglement}. Bohm--Aharonov effect, Berry phase, EPR paradox,
Wheeler's delayed choice experiment or quantum teleportation being
the most paramount examples of the aforementioned. Indeed, one can
go even so far as to claim that because the whole Universe is
inherently quantum correlated one should refrain from using
R\'{e}nyi's entropy altogether. Whether or not these ideas are
viable and whether or not the affiliated entropy is connected with
THC entropy remains yet to be seen.

\vspace{3mm}

As we have shown R\'{e}nyi's entropy has a build--in
predisposition to account for self--similar systems and so it
naturally aspires to be an effective tool to describe phase
transitions (both in equilibrium and non--equilibrium). It is thus
a challenging task to find some connection with such typical tools
of critical phenomena physics as are conformal and renormalization
groups. The latter could in turn bring about a better
understanding of the role of $\alpha$ parameter for systems away
from equilibrium. An interesting application of the former
observation is in the cosmic string physics. In cosmology, unified
gauge theories of particle interactions allow for a sequence of
phase transitions in the very early universe some of which may
lead to defect formation via the so called Kibble--Zurek
mechanism\cite{KiZu1}. Cosmic strings as the most pronounced
example  of such defects, could have important relevance on the
large scale structure formation of the universe or on cosmic
microwave background radiation anisotropies. In astrophysics, for
instance, cosmic strings could play an important r\^{o}le in
dynamics of neutron stars and in the galaxy astrophysics. In usual
cases when the grand--canonical approach is applied it is argued
that at the critical (phase transition) temperature at which
strings tend to fragment into smallest allowed loops,
while large loops become exponentially suppressed - i.e., at
Hagedorn temperature\cite{Hag1}, the correspondence between the
canonical and micro--canonical ensembles breaks down as the
grand--canonical partition function  diverges\cite{Tu2}.  Various
viewpoints with different remedies were lately proposed in the
literature. It seems, however, that non of the treatments has
accommodated the well known fact that the string state--space
acquires approximately self--similar structure which is exact at
critical temperature\cite{Hag1,Tu2}. From this standpoint
R{\'e}ny's statistics appears to be particularly suitable for
generalization of the Hagedorn theory as it could better grasp the
vital features near the critical point. In addition, R{\'e}nyi's
theory can be applied to construct the generalized
grand--canonical partition function for the string network. Our
current results suggest that the new phase transition temperature
should be lower than the one predicted by Hagedorn's theory. It
would be definitely interesting to exploit this further and
contrast our way with the more customary conformal theory
approach. Work along those lines is presently in
progress\cite{JiAr1}.

\vspace{3mm}

Let us finally mention that because symmetry breaking phase
transitions with string--like defects occur in a variety of
physical systems ranging from $~^3$He and $~^4$He superfluids to
the early Universe, with superconductors and liquid crystals in
between, one can hope that predictions based on R\'{e}nyi's
entropy could be directly tested in laboratory. In this
connection, the analysis of vortex tangle \cite{Feynman55}
(turbulence of vortex loops in superfluid phase of $~^4$He) is one
such particularly promising systems with the room--size
experimental setting, (see e.g., \cite{Davis00}).

\section*{Acknowledgements}

P.J. is grateful to Profs.~Larry~S.~Schulman and Sumiyoshi~Abe for
illuminating discussions. We would like to appreciate a partial
support from the ESF network ``COSLAB'' as well as the foreign
professor's grant (P.J.) of the University of Tsukuba.

\section*{Appendix A}

In this appendix we present an alternative way of finding the
unique class of the Kolmogorov--Nagumo functions. Let us start
with Eq.(\ref{fund1}) which we rewrite in the form
\begin{equation}
f(\zeta x) = a(x)f((\zeta -1)x) + f(x)\, , \label{sc22}
\end{equation}
\noi with $\zeta$ being an arbitrary real constant ($\zeta \geq
0$). The latter is equivalent to the equation
\begin{equation}
f(\zeta x) = \frac{1-a^{\zeta}(x)}{1-a(x)} f(x)\, . \label{sc23}
\end{equation}
\noi Note that when $\zeta \rightarrow 0$ then $f(0)= 0$. The
latter should be imposed as a boundary condition on prospective
solutions. The solution of the functional equation (\ref{sc23})
can be easily found, indeed realizing that functions fulfilling
the scaling condition (\ref{sc23}) obey the Euler--type equation
\begin{equation}
x\frac{\partial}{\partial x} f(x) = - \frac{a(x) \ln a(x)}{1-a(x)}
f(x)\, ,
\end{equation}
\noi we may directly write that
\begin{equation}
f(x) = \gamma \exp\left(- \int dx \ \frac{a(x)\ln a(x)}{x
(1-a(x))} \right)\, . \label{cs24}
\end{equation}
\noi Shortly we will see that function (\ref{cs24}) is the only
one fulfilling the functional equation (\ref{sc22}). Let us,
however, first determine the function $a(x)$.  From (\ref{sc22})
follows that
\begin{equation}
a(x) = \frac{f(\zeta x ) - f(x)}{ f((\zeta - 1)x)}\, . \label{a12}
\end{equation}
\noi Because the latter should be true for any $\zeta \geq 0$ we
may safely assume that $\zeta = 1 + \varepsilon/ x$ with
$\varepsilon$ being an infinitesimal. Then with a help of 'l
Hospital rule we obtain
\begin{equation}
a(x) = \frac{f'(x)}{f'(0)}\, , \; \; \Rightarrow f(x) = f'(0)
\int_0^x dy \ a(y) \, . \label{inte3}
\end{equation}
\noi Note that $a(0) =1$. On the other hand (\ref{a12}) may be
equivalently written as
\begin{equation}
a((\zeta -1)x) = \frac{f(\zeta x) - f((\zeta -1)x)}{f(x)}\, .
\end{equation}
\noi Taking now derivative $\partial/\partial(\zeta -1)$, using
(\ref{sc22}) and setting successively $\zeta = 2$ we get
\begin{eqnarray}
a'(x) &=& (a(x) -1)\left( \ln f(x) \right)' =  \frac{a(x)\ln
a(x)}{x}\, , \nonumber \\ &\Rightarrow& \ln a(x) = c x \, .
\label{ln2}
\end{eqnarray}
\noi If the integration constant $c \neq 0$ then $a(x) = \exp(cx)$
and hence (see (\ref{cs24}) and (\ref{inte3}))
\begin{equation}
f(x) = \gamma (\exp(cx) - 1)\, .
\end{equation}
\noi In the latter the condition $f(0) = 0$ was used. We have
defined that $\gamma = f'(0)/c $. In case that $c=0$, we have from
(\ref{ln2}) that $a(x) = const. = 1$ and so
\begin{equation}
f(x) = f'(0)x \, .
\end{equation}
\noi So we see that the compatible Kolmogorov--Nagumo functions
are only linear and exponential ones. We should also note that the
linear $f(x)$ is retrieved from the exponential $f(x)$ in the
limit $c \rightarrow 0$.

\vspace{3mm}

Let us now turn to the point of uniqueness of $f(x)$. For that
purpose let us assume that there are two different functions
$f_1(x)$ and $f_2(x)$ both fulfilling the equation (\ref{sc22})
with an identical $a(x)$ and arbitrary $\zeta \geq 0$, i.e.,
\begin{eqnarray}
f_1(\zeta x) &=& a(x)f_1((\zeta -1)x) + f_1(x)\, , \nonumber \\
f_2(\zeta x) &=&a(x)f_2((\zeta -1)x) + f_2(x)\, .
\end{eqnarray}
\noi Because the latter should hold for any $\zeta \geq 0$ the
following must be true
\begin{eqnarray}
a'(x) &=& (a(x) -1)\left( \ln f_1(x)\right)'\nonumber \\
&=& (a(x) - 1)\left( \ln f_2(x) \right)'\, .
\end{eqnarray}
\noi As a result we have that $\left( \ln f_1(x)\right)' = \left(
\ln f_2(x)\right)'$ and so $f_1(x) = const.\times f_2(x)$, which
confirms that only linear and exponential functions are compatible
with the additivity of information.

\section*{Appendix B}

Here we present a proof that the five postulates of Section IIA
determine uniquely both Shannon's and R\'{e}nyi's entropies. Our
proof consists of four steps:

\vspace{3mm}

$a)$~ Let us denote first ${\mathcal{I}}(1/n, \ldots, 1/n) =
{\mathcal{L}}(n)$. Then from the second and fifth axiom follows
that
\begin{eqnarray}
{\mathcal{L}}(n) &=& {\mathcal{I}}(1/n, \ldots, 1/n, 0)\nonumber
\\
&\leq& {\mathcal{I}}(1/(n+1), \ldots, 1/(n+1)) =
{\mathcal{L}}(n+1)\, ,
\end{eqnarray}
\noindent i.e., ${\mathcal{L}}$ is a non--decreasing function.

\vspace{3mm}

$b)$~To find the explicit form of ${\mathcal{L}}$ we employ the
third postulate. For this purpose we will assume that we have $m$
mutually independent experiments ${\mathcal{A}}^{(1)}, \ldots ,
{\mathcal{A}}^{(m)}$ each with $r$ equally probable outcomes, so
\begin{eqnarray}
{\mathcal{I}}({\mathcal{A}}^{(k)}) = {\mathcal{I}}(1/r, \ldots,
1/r )={\mathcal{L}}(r)\, , \;\;\;\;\; (1\leq k \leq m)\, .
\end{eqnarray}
\noindent Because experiments are independent
${\mathcal{I}}({\mathcal{A}}^{(k)}|{\mathcal{A}}^{(l)} =
{\mathcal{A}}^{(l)}_i) = {\mathcal{I}}({\mathcal{A}}^{(k)})$ for
$k \neq l $ and $\forall i$, axiom 3 (generalized to the case of
$m$ experiments) implies that
\begin{eqnarray}
{\mathcal{I}}({\mathcal{A}}^{(1)}\cap{\mathcal{A}}^{(2)}\cap
\ldots \cap {\mathcal{A}}^{(m)}) &=& \sum_{k=1}^m
{\mathcal{I}}({\mathcal{A}}^{(k)})\nonumber \\
&=& m {\mathcal{L}}(r)\, .
\end{eqnarray}
\noindent On the other hand, the experiment
${\mathcal{A}}^{(1)}\cap{\mathcal{A}}^{(2)}\cap \ldots \cap
{\mathcal{A}}^{(m)}$ consists of $r^m$ equally probable outcomes,
and so
\begin{equation}
{\mathcal{L}}(r^m) = m {\mathcal{L}}(r) \, . \label{fun23}
\end{equation}
\noindent This is nothing but Cauchy's functional
equation\cite{Acz1}. It might be shown\cite{Acz1,Kh1} that for
non--decreasing functions (\ref{fun23}) has a unique solution;
${\mathcal{L}}(r) = \kappa \ln(r)$. The constant $\kappa$ can be
determined from the axiom 2 which then directly implies that
${\mathcal{L}}(r) = \log_2(r)$.

\vspace{3mm}

$c)$~We now determine ${\mathcal{I}}({\mathcal{P}})$ using the
axiom 3. To this extent we will assume that the experiment
${\mathcal{A}} = ({\mathcal{A}}_1,{\mathcal{A}}_2, \dots,
{\mathcal{A}}_n)$ is described by the distribution ${\mathcal{P}}
=\{p_1,p_2, \ldots, p_n\}$ with $p_k \, (1\leq k \leq n)$ being
rational numbers, say
\begin{equation}
p_k = \frac{g_k}{g}\, , \;\;\;\; \sum_{k=1}^n g_k = g\, , \;\;\;\;
g_k \in \mathbb{N}\, .
\end{equation}
\noindent Let us have further an experiment ${\mathcal{B}} =
({\mathcal{B}}_1, {\mathcal{B}}_2,\ldots, {\mathcal{B}}_g)$ and
let ${\mathcal{Q}} = \{q_1, q_2, \ldots, q_g \}$ is the associated
distribution. We split $({\mathcal{B}}_1,{\mathcal{B}}_2,\ldots,
{\mathcal{B}}_g)$ into $n$ groups containing $g_1, g_2, \ldots,
g_n$ events respectively. Consider now a particular situation in
which whenever event ${\mathcal{A}}_i$ in ${\mathcal{A}}$ happens
then in ${\mathcal{B}}$ all the $g_k$ events of $k$-th group occur
with the equal probability $1/g_k$ an all the other events in
${\mathcal{B}}$ have probability zero. Hence
\begin{equation}
{\mathcal{I}}({\cal{B}}|{\cal{A}} = {\cal{A}}_k) =
{\cal{I}}(1/g_k, \ldots, 1/g_k) = \log_2 g_k\,,
\end{equation}
\noindent and so
\begin{equation}
{\mathcal{I}}({\cal{B}}|{\cal{A}}) = f^{-1}\left( \sum_{k=1}^n
\varrho_k(\alpha) f(\log_2 g_k) \right)\, . \label{cond1}
\end{equation}
\noi On the other hand, ${\cal{I}}({\cal{A}}\cap {\cal{B}})$ can
be directly evaluated. Realizing that the joint probability
distribution corresponding to ${\cal{A}}\cap {\cal{B}}$ is
\begin{eqnarray}
{\cal{R}} &=& \left\{ r_{kl} = p_k q_{l|k}
 \right\}\nonumber \\
 &&\nonumber \\
&=& \{ \underbrace{\frac{p_1}{g_1}, \ldots,  \frac{p_1}{g_1},
}_{g_1 \times} \underbrace{\frac{p_2}{g_2},
\ldots,\frac{p_2}{g_2}, }_{g_2 \times} \ldots,
\underbrace{\frac{p_n}{g_n}, \ldots, \frac{p_n}{g_n} }_{g_n
\times}\}
\nonumber \\
&=& \left\{ 1/g, \ldots, 1/g  \right\}\, ,
\end{eqnarray}
\noi we obtain that ${\mathcal{I}}({\mathcal{A}}\cap {\mathcal{B}}) =
{\mathcal{L}}(g) = \log_2 g$. Applying the axiom 3 then
\begin{eqnarray}
{\mathcal{I}}({\mathcal{P}}) &=& \log_2 g - f^{-1}\left(
\sum_{k=1}^n \varrho_k(\alpha) f(\log_2 g_k))\right)\nonumber \\
&=& \log_2 g - f^{-1} \left( \sum_{k=1}^n \varrho_k(\alpha)
f(\log_2p_k + \log_2 g) \right)\nonumber \\
&=& {\mathcal{L}}(g) -f^{-1} \left( \sum_{k=1}^n \varrho_k(\alpha)
f(\log_2p_k + {\mathcal{L}}(g)) \right)\, .
\end{eqnarray}
\noi Let us define $f_y(x) = f(-x-y)$ ($\Rightarrow f^{-1}(x) + y
= -f^{-1}_y(x)$). Then
\begin{equation}
{\mathcal{I}}({\mathcal{P}}) = f^{-1}_{{\mathcal{L}}(g)} \left(
\sum_{k=1}^{n} \varrho_k(\alpha)
f_{{\mathcal{L}}(g)}({\mathcal{I}}_k) \right)\, . \label{lg1}
\end{equation}
\noi By axiom 4 $f(x)$ is invertible in $[0, \infty)$ and so both
$f_{{\cal{L}}(g)}$ and $f_{{\cal{L}}(g)}^{-1}$ are continuous on
$[0,\infty)$. Applying now the postulate 1 (axiom of continuity)
we may extend the result (\ref{lg1}) from rational $p_k$'s to any
real valued $p_k$'s defined in [0,1].

\vspace{3mm}

Let us consider now the case of independent events (i.e.,
${\mathcal{I}}({\mathcal{B}}|{\mathcal{A}}) =
{\mathcal{I}}({\mathcal{B}})$). From Section IIA (and/or Appendix
A)  we already know that in this case the only candidate for
$f_{{\mathcal{L}}(g)}$ is a linear function or a linear function
of an exponential function.  Bearing in mind that two functions
which are linear functions of each other give the same mean (see
Section IIA) we may choose either $f_{{\mathcal{L}}_(g)}(x) = x$
or $f_{{\mathcal{L}}_(g)}(x) = 2^{(\lambda -1)x}, \, \lambda \neq
1$. Consequently from (\ref{lg1}) we may write
\begin{eqnarray}
{\mathcal{I}}({\mathcal{P}}) &=& \frac{1}{(\lambda -1)} \log_2
\left( \sum_{k=1}^n p_k^{\alpha - \lambda + 1}\right)\nonumber \\
&+& \frac{1}{(1- \lambda)} \log_2 \left( \sum_{k=1}^n p_k^\alpha
\right)\, . \label{rensh1}
\end{eqnarray}
\noi  It should be also noticed that from the axiom 5 follows that
$(\alpha - \lambda +1 ) > 0$ and $\alpha > 0$. Within the scope of
previous inequalities Eq.(\ref{rensh1}) is valid for any
$\lambda$. It should be particularly noticed that
${\mathcal{I}}({\mathcal{P}})$ is continuous  at  $\lambda  =1$ as
both  the left and right limit coincide.   It   can   be easily
checked  that  $\lambda  =1$ corresponds precisely to the case of
$f_{{\mathcal{L}}_(g)}(x) = x$. Quantity (\ref{rensh1}) was
firstly proposed by Kapur\cite{Kapur} and  named  the {\em entropy
of order $2-\lambda$ and type $\alpha$}.

\vspace{3mm}

Finally, it should be born in mind that because the mean
(\ref{cond1}) is unchanged under linear transformation of function
$f(x)$ we could, from the very beginning, restrict ourselves to
only {\em positive} invertible functions on $[0, \infty)$.

\vspace{3mm}

$d)$~In the last step we will specify the relationship between
$\alpha$ an $\lambda$. Using the fact that the experiment
${\mathcal{A}}\cap{\mathcal{B}}$ has the (joint) probability
distribution ${\mathcal{R}} = \{r_{kl} = p_k q_{l|k} \}$ we have
\begin{eqnarray} {\mathcal{I}}({\mathcal{A}}\cap{\mathcal{B}})&=&
\frac{1}{(\lambda -1)} \log_2
\left( \sum_{k,l} (p_k q_{l|k})^{\alpha - \lambda + 1}\right)\nonumber \\
&+& \frac{1}{(1- \lambda)} \log_2 \left( \sum_{k,l}(p_k
q_{l|k})^\alpha \right)\, , \label{A|B0}
\end{eqnarray}
\noi and
\begin{eqnarray}
{\mathcal{I}}({\mathcal{B}}|{\mathcal{A}})&=& \frac{1}{(\lambda
-1)} \log_2 \left( \sum_{k} p_k ^\alpha \right)\nonumber
\\
&+& \frac{1}{(1-\lambda)} \log_2 \left( \sum_{k} p_k^{\alpha}
\frac{\sum_{l} (q_{l|k})^{\alpha}}{\sum_{l} (q_{l|k})^{\alpha -
\lambda + 1}}\right)\, . \label{A|B}
\end{eqnarray}
\noi Eq.(\ref{A|B}) is a result of the fact that
\begin{displaymath}
2^{(1-\lambda){\mathcal{I}}({\mathcal{B}}|{\mathcal{A}}={\mathcal{A}}_k)}
= \frac{\sum_{l} (q_{l|k})^{\alpha}}{\sum_{l} (q_{l|k})^{\alpha -
\lambda +1}}\, ,
\end{displaymath}
\noi and that $f_{{\mathcal{L}}(g)}(x) = 2^{(\lambda -1)x}
\Rightarrow f(x) = 2^{(1-\lambda)x}$. Combining the axiom 3 and
Eqs.(\ref{A|B0})--(\ref{A|B}) we obtain for $\lambda \neq 1$ the
identity
\begin{eqnarray}
&&\frac{\sum_{k} p_k^{\alpha - \lambda +1} \sum_l
(q_{l|k})^{\alpha -\lambda+1 }}{\sum_{k} p_k^{\alpha} \sum_l
(q_{l|k})^{\alpha}}\nonumber \\
&&\mbox{\hspace{25mm}} = \frac{\sum_k p_k^{\alpha - \lambda +
1}}{\sum_k p_k^{\alpha} \frac{\sum_l (q_{l|k})^{\alpha}}{\sum_l
(q_{l|k})^{\alpha - \lambda + 1}}}\, . \label{rel1}
\end{eqnarray}
\noindent Introducing the random variable
\begin{displaymath}
{\mathcal{Q}}^{(\alpha\lambda)} = \{\sum_l (q_{l|k})^{\alpha
-\lambda +1} \}\, ,
\end{displaymath}
\noindent we may equivalently rewrite (\ref{rel1}) as
\begin{eqnarray}
&&\frac{\sum_{k,l}r_{kl}^{\alpha - \lambda + 1}   }{\sum_{k,l}
r_{kl}^{\alpha}} = \frac{\sum_{k,l} r_{kl}^{\alpha - \lambda +
1}/{\mathcal{Q}}_k^{(\alpha \lambda)} }{\sum_{k,l} r_{kl}^{\alpha
}/{\mathcal{Q}}_k^{(\alpha \lambda)}}\, ,\nonumber \\
&&\nonumber \\
&\Leftrightarrow & \; \langle 1/{\mathcal{Q}^{(\alpha \lambda)}}
\rangle_{\alpha} = \langle 1/{\mathcal{Q}}^{(\alpha \lambda)}
\rangle_{\alpha -\lambda + 1} \, . \label{a5}\end{eqnarray}
\noi Here $\left\langle \ldots \right\rangle_{x} $ is defined with
respect to the distribution
\begin{displaymath}
{\mathcal{P}} = \{ \sum_l (r_{kl})^{x}/ \sum_{k,l} (r_{kl})^{x}
\}\, .
\end{displaymath}
\noi Because $p_k$'s are arbitrary, equality (\ref{a5}) happens if
and only if ${\mathcal{Q}}^{(\alpha\lambda)}$ is a
constant\cite{Hardy1}. The latter implies that
\begin{equation}
\sum_l (q_{l|k})^{\alpha - \lambda + 1} = const.\, , \;\;\;\;\;
\mbox{for}\;\, \forall k \;\, \mbox{and} \;\, \forall q_{l|k}\, .
\label{jen23}
\end{equation}
\noi It is easy to see that Eq.(\ref{jen23}) is satisfied only when $\alpha =
\lambda$. Substituting $\lambda = \alpha$ into (\ref{rensh1}) we find
\begin{equation} {\mathcal{I}}({\mathcal{P}}) =
{\mathcal{I}}({\mathcal{A}}) = \frac{1}{1-\alpha} \log_2 \sum_k
(p_k)^{\alpha}\, .
\end{equation}
\noi The proof for $\lambda =1$ follows the analogous route. This
proves our assertion.

\section*{Appendix C}\label{ApC}

In this appendix we derive some basic properties of the
information measure
${\mathcal{I}}_{\alpha}({\mathcal{B}}|{\mathcal{A}})$.

\vspace{3mm}

From Appendix B we know that $f(x)$ compatible with axioms 1--5 is
(up to a linear combination) either $x$ or $2^{(1-\alpha)x}$. Then
${\mathcal{I}}({\mathcal{B}}|{\mathcal{A}})$ appearing in the
axiom 3 turns out to have the form
\begin{equation}
{\mathcal{I}}_{\alpha}({\mathcal{B}}|{\mathcal{A}}) =
\frac{1}{(1-\alpha)} \log_2
\left(\frac{\sum_{k,l}(r_{kl})^{\alpha}}{\sum_{k}p_k^{\alpha}}\right)\,
, \label{SS00}\end{equation}
\noi with ${\mathcal{P}}({\mathcal{A}} \cap {\mathcal{B}}) =
\{r_{kl}= p_k q_{l|k} = q_l p_{k|l}$\}. We have reintroduced the
sub--index $\alpha$ to emphasize the parametric dependence of
${\mathcal{I}}$. It results from (\ref{SS00}) that for every
$\alpha$
\begin{equation}
0\leq {\mathcal{I}}_{\alpha}({\mathcal{B}}|{\mathcal{A}}) \leq
\log_2 n\, , \label{SS0}\end{equation}
\noi where $n$ is the number of outcomes in the experiment
${\mathcal{B}}$. Indeed, $0\leq
{\mathcal{I}}_{\alpha}({\mathcal{B}}|{\mathcal{A}})$ holds due to
a simple fact that for a fixed $k$ and $\alpha > 1$
\begin{equation}
\sum_{l}(r_{kl})^{\alpha} = p_k^{\alpha} \ \sum_l
(q_{l|k})^{\alpha} \leq p_k^{\alpha}\, , \label{SS1}
\end{equation}
\noi (realize that $\sum_l q_{l|k} =1 $). Equality in (\ref{SS1})
is clearly valid if and only if for any $k$ there exists just one
$l = l(k)$ such that $q_{l(k)|k} = 1$ and $0$ otherwise. The
latter means that outcomes of ${\mathcal{A}}$ uniquely determine
outcomes of ${\mathcal{B}}$ and hence we do not learn any new
information about ${\mathcal{B}}$ by knowing ${\mathcal{A}}$. In
such a case (\ref{SS00}) gives
${\mathcal{I}}_{\alpha}({\mathcal{B}}|{\mathcal{A}}) = 0$. This is
what one would naturally expect from a conditional information.

%
%
%
%
%
%
%
%
%

\vspace{3mm}

Similarly, for $0<\alpha<1$ the reverse inequality in (\ref{SS1})
holds and hence $\sum_l (r_{kl})^{\alpha} \geq p_k^{\alpha}$
(former comments about the equality apply here as well). This
proves our assertion about the LHS inequality in (\ref{SS0}).

\vspace{3mm}

On the other hand, the RHS inequality in Eq.(\ref{SS0}) holds
because for $\alpha > 1$, $\sum_l (q_{l|k})^{\alpha}$ is a convex
function which has its minimum at $q_{l|k}=1/n$ (for $\forall \
l,k$). So
\begin{displaymath}
\sum_l (q_{l|k})^{\alpha} \geq n^{1-\alpha}\, ,
\end{displaymath}
\noi while for $0<\alpha<1$ the opposite inequality holds. Thus
\begin{eqnarray}
 {\mathcal{I}}_{\alpha}({\mathcal{B}}|{\mathcal{A}}) &=&
\frac{1}{(1-\alpha)}\log_2 \left( \frac{\sum_k p_k^{\alpha} \sum_l
(q_{l|k})^{\alpha}}{\sum_k p_k^{\alpha}} \right)\nonumber \\
&\leq& \log_2 n\, . \label{SS5}
\end{eqnarray}
%
\noi Inequality (\ref{SS5}) may be viewed as a weak version of the
well known  $\alpha = 1$ case where
${\mathcal{H}}({\mathcal{B}}|{\mathcal{A}}) \leq
{\mathcal{H}}({\mathcal{B}})$ with equality if and only if
${\mathcal{B}}$ and ${\mathcal{A}}$ are independent
experiments\cite{Kh1} (i.e., knowing outcomes of ${\mathcal{A}}$
does not have any effect on the distribution of outcomes of
${\mathcal{B}}$). However, aforesaid does not generally hold for
$\alpha \neq 1$. This is because
\begin{eqnarray}
{\mathcal{I}}_{\alpha}({\mathcal{B}}) -
{\mathcal{I}}_{\alpha}({\mathcal{B}}|{\mathcal{A}}) =
\frac{1}{(1-\alpha)} \log_2 \left( \frac{\sum_{l,k}p_k^{\alpha}
q_{l}^{\alpha} }{\sum_{l,k}(p_k q_{l|k})^{\alpha}  } \right)\! ,
\label{SS6}
\end{eqnarray}
\noi and the identity
\begin{equation}
\mbox{\huge{(}}\sum_{k,l}(p_k \,
q_l)^{\alpha}\mbox{\huge{)}}^{1/(1-\alpha)} =
\mbox{\huge{(}}\sum_{k,l}(p_k \,
q_{l|k})^{\alpha}\mbox{\huge{)}}^{\!\! 1/(1-\alpha)}\, ,
\label{SS7}\end{equation}
\noi can be fulfilled for $\alpha\neq 1$ in numerous
ways\cite{Mar1} without assuming that $q_{l|k} = q_k$  (for
example, in the $\alpha =2 $ case we may chose; ${\mathcal{P}} =
\{1/n\}, {\mathcal{Q}} = \{1/n\}$ and
${\mathcal{P}}({\mathcal{B}}|{\mathcal{A}}) = \{1, 0,0, \ldots,
0\}$). However, in the limiting case $\alpha \rightarrow 1$
Eq.(\ref{SS7}) turn out to be
\begin{equation}
2^{-\sum_{k,l}p_kq_l \log_2(p_k q_l)} = 2^{-\sum_{k,l} r_{kl}
\log_2(r_{kl})}\, ,
\end{equation}
\noi which has the solution if and only if $q_{l|k} =q_l$, i.e.,
in the case of independent events\cite{Kh1}. Yet still,
${\mathcal{I}}_{\alpha}({\mathcal{B}}|{\mathcal{A}})$, $\alpha\neq
1$ can be, in a sense, viewed as conditional information. This is
so because when ${\mathcal{B}}$ and ${\mathcal{A}}$ are
independent then from (\ref{SS6}) follows that
${\mathcal{I}}_{\alpha}({\mathcal{B}}) =
{\mathcal{I}}_{\alpha}({\mathcal{B}}|{\mathcal{A}})$.  Opposite
implication, as we have seen, is not valid in general. The
opposite implication is, however, valid when ${\mathcal{B}}$ has
an equiprobable distribution. The latter is a simple consequence
of Jensen's inequality because for $\alpha > 1$
\begin{displaymath}
p_k^{\alpha} = \left(\sum_l q_l \frac{r_{kl}}{q_l}
\right)^{\!\!\alpha} \leq \sum_l q_l \left(\frac{r_{kl}}{q_l}
\right)^{\!\!\alpha} = \sum_l q_l (p_{k|l})^{\alpha}\, ,
\end{displaymath}
\noi and so for ${\mathcal{P}}({\mathcal{B}})= {\mathcal{Q}}= \{
q_l = 1/n \}$
\begin{eqnarray}
&& \frac{\sum_{l,k} q_l^{\alpha}
p_k^{\alpha}}{\sum_{k,l}(r_{kl})^{\alpha}} \leq \frac{\sum_l
q^{\alpha}_l \sum_{j,k} q_j (p_{k|j})^{\alpha}}{\sum_{l,k}
q_l^{\alpha} (p_{k|l})^{\alpha}} = 1\, ,\nonumber \\
&&\nonumber \\ &&\Rightarrow \;
{\mathcal{I}}_{\alpha}({\mathcal{B}}) -
{\mathcal{I}}_{\alpha}({\mathcal{B}}|{\mathcal{A}}) \geq 0\, ,
\end{eqnarray}
\noi with equality if and only if the equality in Jensen's
inequality holds. This happens only when $p_{k|l}$ is a constant
for $\forall \, l$, i.e., when ${\mathcal{A}}$ and ${\mathcal{B}}$
are independent. Counterpart with $0 < \alpha < 1$ can be proved
in exactly the same way.

%
\section*{Appendix D}

In this appendix we derive relations (\ref{ren3}) and
(\ref{ren4}). We begin with the notion of the integration of
continuous functions defined on fractal sets\cite{Edgar,Barn}.
Consider a fractal set $M$ embedded in a $d$--dimensional space.
Let us cover the set with a mesh $M^{(l)}$ of $d$--dimensional
(disjoint) cubes $M_i^{(l)}$ of size $l^d$ and let $N_l(M)$ is a
minimal number of the cubes needed for the covering. Functions
with the support in the mesh are called {\em simple} if they can
be decomposed in the following way:
\begin{equation} {\mathcal{G}}^{(l)}({\mathbf{x}}) =
\sum_{i}^{N_l}{\mathcal{G}}^{(l)}_i \chi_{i}^{(l)}({\mathbf{x}})\,
. \label{b1}
\end{equation}
\noi Here $\chi_{i}^{(l)}$ are characteristic functions, i.e.
\begin{eqnarray}
\chi_{i}^{(l)} = \left\{
\begin{array}{l}
 1 \; \mbox{if} \;
{\mathbf{x}}\in M^{(l)}_i\\
 0 \; \mbox{if} \; {\mathbf{x}}\not\in M^{(l)}_i\, .
\end{array} \right.
\end{eqnarray}
\noi Then the integral $\int_M d\mu \ {\mathcal{G}}^{(l)}$ is
defined as
\begin{equation}
\int_M d\mu({\bf{x}}) \ {\mathcal{G}}^{(l)}({\mathbf{x}}) =
\sum_{i}^{N_l} {\mathcal{G}}^{(l)}_i \ \mu^{(l)} (M_i^{(l)})\, ,
\end{equation}
\noi where the measure $\mu^{(l)}$ is the measure on the covering
mesh. The precise form of the measure will be specified shortly.
On the covering mesh $M^{(l)}$ we can build a $\sigma$-structure
in a usual way. As a result,  if $\mathcal{G}$ is a nonnegative
$\mu^{(l)}$ measurable function then ${\mathcal{G}}({\mathbf{x}})
= \lim_{l\rightarrow 0}{\mathcal{G}}^{(l)}({\mathbf{x}})$ for all
${\mathbf{x}}\in M^{(l)}$, for some sequence $\{
{\mathcal{G}}^{(l)}_i\}$ of monotonic increasing nonnegative
simple functions. Owing to this fact we may define
\begin{equation}
\int_M d\mu({\mathbf{x}}) \ {\mathcal{G}}({\mathbf{x}}) =
\lim_{l\rightarrow 0} \sum_{i=1}^{N_l}{\mathcal{G}}^{(l)}_i \
\mu^{(l)} (M_i^{(l)})\, . \label{b3}
\end{equation}
\noi In this connection it is important to notice that due to the
scaling prescription (\ref{fra1})
\begin{equation}
\log l^D = - \log N_l + o(l^0) \; \; \Rightarrow \; \; l^D N_l =
V_{l} \rightarrow V\, .
\end{equation}
\noi Here $V_l$ is the pre--fractal volume which in the small $l$
limit converges to the true fractal volume $V$. Natural candidate
for $\mu^{(l)} (M_i^{(l)})$ is the fraction $V(M^{(l)})/N_l$ which
in the small $l$ limit behaves as\footnote{It should be noted that
the measure just defined basically coincides with the
$D$--dimensional Hausdorff measure.}: $l^D = n^{-D}$. So
particularly when ${\mathcal{F}}$ is a continuous PDF we have
\begin{equation}
\int_{M} d\mu({\mathbf{x}}) \ {\mathcal{F}}({\mathbf{x}}) =
\lim_{l\rightarrow 0} \sum_{i=1}^{N_l} {\mathcal{F}}_i^{(l)} \ l^D
\, . \label{b4}
\end{equation}
\noi The integrated probability of the $k$--th cube is thus
$p_{nk} = {\mathcal{F}}_{k}^{(l)}l^D$. A simple consistency check
can be demonstrated on $p_{nk} = {\mathcal{E}}_{nk}$. Indeed, from
Section IV~A we know that ${\mathcal{E}}_{nk} = l^D/V_l$ and so
may write
\begin{eqnarray}
1 &=& \lim_{l \rightarrow 0} \sum_{k=1}^{N_l} {\mathcal{E}}_{nk} =
\lim_{l \rightarrow 0} \sum_{k=1}^{N_l} \frac{l^D}{V_l}
 = \int_{M}
d\mu \ \frac{1}{V} = 1\, .
\end{eqnarray}
\noi We thus see that the integral prescription (\ref{b4}) applies
correctly in the case of uniform distributions.

\vspace{3mm}

Using now the renormalization prescription (\ref{ren3})
\begin{eqnarray}
\tilde{{\mathcal{I}}}_{\alpha}({\mathcal{F}}) &\equiv& \lim_{l
\rightarrow 0} \left( {\mathcal{I}}_{\alpha}({\mathcal{P}}_n) -
{\mathcal{I}}_{\alpha}({\mathcal{E}}_n) \right)  \nonumber \\
&=& \lim_{l \rightarrow 0} \left(\frac{1}{1-\alpha} \log_2 \left(
\sum_{i=1}^{N_l} \left({\mathcal{F}}^{(l)}_i l^D\right)^{\alpha} /
\sum_{i=1}^{N_l} \frac{l^{D\alpha}}{V^{\alpha}_l}
\right) \right)\nonumber \\
&=& \lim_{l \rightarrow 0} \left(\frac{1}{1-\alpha} \log_2 \left(
\sum_{i=1}^{N_l} \left({\mathcal{F}}^{(l)}_i \right)^{\alpha} l^D
/ \sum_{i=1}^{N_l} \frac{l^{D}}{V^{\alpha}_l}
\right) \right)\nonumber \\
&=& \frac{1}{1-\alpha} \log_2 \left( \frac{\int_M d\mu \
{\mathcal{F}}^{\alpha}}{\int_M d\mu \ 1/V^{\alpha} } \right)\, .
\label{b5}
\end{eqnarray}
\noi If we use the renormalization prescription (\ref{ren4}) (or
equivalently when we set $V=1$ for
${\mathcal{I}}_{\alpha}({\mathcal{E}}_n)$ in (\ref{b5})) we easily
see that
\begin{eqnarray}
{{\mathcal{I}}}_{\alpha}({\mathcal{F}}) &\equiv& \lim_{l
\rightarrow 0} \left( {\mathcal{I}}_{\alpha}({\mathcal{P}}_n) -
{\mathcal{I}}_{\alpha}({\mathcal{E}}_n)|_{V=1} \right)  \nonumber \\
&=& \lim_{l \rightarrow 0} \left(
{\mathcal{I}}_{\alpha}({\mathcal{P}}_n)  + D \log_2 l \right)  \nonumber \\
&=& \frac{1}{1-\alpha} \log_2 \left( \int_M d\mu \
{\mathcal{F}}^{\alpha}  \right)\, . \label{b6}
\end{eqnarray}
\noi Our renormalization prescription is obviously consistent only
when integrals on the RHS of (\ref{b5}) and (\ref{b6}) exist.

\section*{Appendix E}

We show here that R{\'e}nyi's entropy
${\cal{I}}_{\alpha}^{\scriptscriptstyle (d)}({\cal{F}})$ is not
invariant under a transformation of the continuous  random
variable ${\cal{A}}^{\scriptscriptstyle (d)}$ while
$\hat{{\cal{I}}}_{\alpha}^{\scriptscriptstyle (d)}({\cal{F}})$ is.
Note first that in a discrete case, outcomes ${\cal{A}}_1, \ldots,
{\cal{A}}_n$ have the same probability distribution $p_1, \ldots,
p_n$ as outcomes $h({\cal{A}}_1), \ldots, h({\cal{A}}_n)$, where
$h(\ldots)$ is an arbitrary ``well behaved" function. Hence
R{\'e}nyi's entropy for such a system is invariant under the
$h$--transformation. However, in the continuous case even the
simplest linear transformation ${\cal{A}}^{\scriptscriptstyle (d)}
\rightarrow c{\cal{A}}^{\scriptscriptstyle (d)}$ does not leave
${\cal{I}}_{\alpha}^{\scriptscriptstyle (d)}({\cal{F}})$
invariant, indeed after rescalling ${\cal{A}}^{\scriptscriptstyle
(d)}$ to $c{\cal{A}}^{\scriptscriptstyle (d)}$ we obtain
\begin{eqnarray*}
(c{\mathcal{A}}^{\scriptscriptstyle (d)})_n &\equiv &
\tilde{{\mathcal{A}}}_n^{\scriptscriptstyle (d)}\nonumber \\ &=&
\left(c\frac{\left[(n c) {\mathcal{A}}_1 \right]}{(nc)},
c\frac{\left[(nc) {\mathcal{A}}_2\right] }{(nc)}, \ldots ,
c\frac{\left[(nc)
{\mathcal{A}}_d \right]}{(nc)} \right)\nonumber \\
 &=& c
{\mathcal{A}}_{(nc)}^{\scriptscriptstyle (d)}\, ,
\end{eqnarray*}
\noindent and so
\begin{eqnarray}
{\cal{I}}_{\alpha}^{\scriptscriptstyle (d)}(c
{\mathcal{A}}^{\scriptscriptstyle (d)}) &=&
\lim_{n\rightarrow\infty}\left(
{\mathcal{I}}_{\alpha}(\tilde{\mathcal{A}}_{n}^{\scriptscriptstyle
(d)}) - d \log_2 n
\right)\nonumber \\
&=& \lim_{(nc)\rightarrow \infty} \left( {\mathcal{I}}_{\alpha}(c
{\mathcal{A}}_{(nc)}^{\scriptscriptstyle (d)}) - d \log_2 (nc) + d
\log_2 c \right)
\nonumber \\
&=&  {\cal{I}}_{\alpha}^{\scriptscriptstyle (d)}
({\cal{A}}^{\scriptscriptstyle (d)}) + d\log_2 c \, .
\label{ren45}
\end{eqnarray}
\noi So ${\cal{I}}_{\alpha}^{\scriptscriptstyle (d)}(c
{\mathcal{A}}^{\scriptscriptstyle (d)}) \neq
{\cal{I}}_{\alpha}^{\scriptscriptstyle (d)}
({\cal{A}}^{\scriptscriptstyle (d)})$. Situation becomes, however,
different when we consider
$\tilde{\mathcal{I}}^{\scriptscriptstyle (d)}_{\alpha}(c
{\mathcal{A}}^{\scriptscriptstyle (d)})$. This is because we can
rewrite  $\tilde{\mathcal{I}}^{\scriptscriptstyle (d)}_{\alpha}(c
{\mathcal{A}}^{\scriptscriptstyle (d)})$ as
\begin{eqnarray}
 \tilde{\mathcal{I}}^{\scriptscriptstyle (d)}_{\alpha}(c
{\mathcal{A}}^{\scriptscriptstyle (d)}) &=&
\lim_{n\rightarrow\infty}\left({\mathcal{I}}_{\alpha}(
\tilde{{\mathcal{A}}}^{\scriptscriptstyle (d)}_n) - d \log_2 n
\right)
\nonumber \\
&-& \lim_{n\rightarrow\infty} \left({\mathcal{I}}_{\alpha}(
{\mathcal{E}}^{\scriptscriptstyle (d)}_{(nc)} ) - d \log_2 n
\right)\, . \label{resc2}
\end{eqnarray}
\noi Here we have used ${\mathcal{E}}^{\scriptscriptstyle
(d)}_{(nc)} $ instead of ${\mathcal{E}}^{\scriptscriptstyle
(d)}_{n}$ because the rescalling changes also the volume $V$ of
the outcome space into $cV$. A simple consequence of
Eq.(\ref{resc2}) is that $\tilde{\mathcal{I}}^{\scriptscriptstyle
(d)}_{\alpha}(c {\mathcal{A}}^{\scriptscriptstyle (d)}) =
\tilde{\mathcal{I}}^{\scriptscriptstyle
(d)}_{\alpha}({\mathcal{A}}^{\scriptscriptstyle (d)})$. In fact,
when $h = (h_1, \ldots, h_d)$ is an invertible and differentiable
(vector) function it is simple to rewrite
$\tilde{\mathcal{I}}^{\scriptscriptstyle
(d)}_{\alpha}({\mathcal{F}})$ in a fully covariant manner. Indeed,
realizing that scalar density transforms as
\begin{equation}
{\mathcal{F}}({\mathbf{x}})  = \left| \frac{\partial
{\mathbf{y}}}{\partial {\mathbf{x}}} \right| \
\hat{\mathcal{F}}({\mathbf{y}})\, .
\end{equation}
\noi (here ${\mathbf{y}} = h({\mathbf{x}})$) we also know that
\begin{equation}
1/V = \left| \frac{\partial {\mathbf{y}}}{\partial {\mathbf{x}}}
\right| \ m({\mathbf{y}})\, ,
\end{equation}
\noi (here $m({\mathbf{y}})$ denotes the $h$--transformed uniform
PDF). Then we see that
\begin{eqnarray}
\tilde{\mathcal{I}}^{\scriptscriptstyle
(d)}_{\alpha}({\mathcal{A}}^{\scriptscriptstyle (d)}) &=&
\frac{1}{(1-\alpha )} \log_2 \left(\int_V d^d {\mathbf{x}} \
{\mathcal{F}}^{\alpha}({\mathbf{x}}) \ V^{\alpha
-1} \right)\nonumber \\
&=& \frac{1}{(1-\alpha )} \log_2 \left(\int_{h(V)} d^d
{\mathbf{y}}
\left(\frac{\hat{\mathcal{F}}({\mathbf{y}})}{m({\mathbf{y}})}\right)^{\!\!\alpha}
\mbox{\hspace{-2mm}}m({\mathbf{y}}) \right)\nonumber \\
&=& \tilde{\mathcal{I}}^{\scriptscriptstyle
(d)}_{\alpha}(h({\mathcal{A}}^{\scriptscriptstyle (d)})) \, .
\end{eqnarray}
\noi If $h_1$ and $h_2$ are any two invertible and differentiable
vector functions so is their composition $h_2\circ h_1$ and then
\begin{eqnarray}
&&\tilde{\mathcal{I}}^{\scriptscriptstyle (d)}_{\alpha}
({\mathcal{A}}^{\scriptscriptstyle(d)}) = \;
\tilde{\mathcal{I}}^{\scriptscriptstyle (d)}_{\alpha}
(h_1({\mathcal{A}}^{\scriptscriptstyle(d)}))\nonumber \\
&&\mbox{\hspace{5mm}}= \frac{1}{(1-\alpha)} \log_2 \left(
\int_{h_1(V)} d^d {\mathbf{y}}
\left(\frac{{\mathcal{F}}_1({\mathbf{y}})}{m_1({\mathbf{y}})}
\right)^{\!\!\alpha}\mbox{\hspace{-1mm}} m_1({\mathbf{y}})\right)\nonumber \\
&&\mbox{\hspace{5mm}}= \frac{1}{(1-\alpha)} \log_2 \left(
\int_{h_2\circ h_1(V)} d^d {\mathbf{z}}
\left(\frac{{\mathcal{F}}_2({\mathbf{z}})}{m_2({\mathbf{z}})}
\right)^{\!\!\alpha}\mbox{\hspace{-1mm}} m_2({\mathbf{z}})\right)\nonumber \\
&&\nonumber \\
 &&\mbox{\hspace{5mm}}= \; \tilde{\mathcal{I}}^{\scriptscriptstyle (d)}_{\alpha}(h_2\circ
h_1({\mathcal{A}}^{\scriptscriptstyle (d)}))\, , \label{rep2}
\end{eqnarray}
\noi with
\begin{eqnarray}
&&{\mathcal{F}}_1({\mathbf{y}}) \left| \frac{\partial
{\mathbf{y}}}{\partial {\mathbf{x}}}\right| =
{\mathcal{F}}({\mathbf{x}})\, , \;\;\;\;\;
{\mathcal{F}}_2({\mathbf{z}}) \left| \frac{\partial
{\mathbf{z}}}{\partial {\mathbf{y}}}\right| =
{\mathcal{F}}_1({\mathbf{y}})\, , \nonumber \\
&&m_1({\mathbf{y}})\left| \frac{\partial {\mathbf{y}}}{\partial
{\mathbf{x}}}\right| = 1/V\, , \;\;\;\;\;\;
m_2({\mathbf{z}})\left| \frac{\partial {\mathbf{z}}}{\partial
{\mathbf{y}}}\right| = m_1({\mathbf{y}})\, , \label{rep4}
\end{eqnarray}
\noi and ${\mathbf{y}} = h_1({\mathbf{x}}), \, {\mathbf{z}} =
h_2({\mathbf{y}}) = h_2\circ h_1 ({\mathbf{x}})$. Thus
$\tilde{\mathcal{I}}^{\scriptscriptstyle
(d)}_{\alpha}({\mathcal{F}})$ is invariant under the
outcome--space reparametrization. In addition, if we restrict our
consideration only to the class of transformations which have also
differentiable inverse i.e., diffeomorphisms, we see from
(\ref{rep2}) and (\ref{rep4}) that the information measure
$\tilde{\mathcal{I}}^{\scriptscriptstyle (d)}_{\alpha}$ is
invariant with respect to the group of diffeomorphisms. This fact
was firstly realized by E.T.~Jaynes in the context of Shannon's
entropy\cite{Jay2}. As a matter of fact, when setting $\alpha =1$
we obtain from (\ref{rep2}) that
\begin{eqnarray}
{\tilde{\mathcal{H}}}({\mathcal{F}}) &=& \lim_{\alpha \rightarrow
1}\frac{1}{(1-\alpha )} \log_2 \left(\int_{h(V)} d^d {\mathbf{y}}
\left(\frac{\hat{\mathcal{F}}({\mathbf{y}})}{m({\mathbf{y}})}\right)^{\!\!\alpha}
\mbox{\hspace{-1mm}} m({\mathbf{y}}) \right)
\nonumber \\
&=& -\int_{h(V)}d^d {\mathbf{y}}\
{\hat{\mathcal{F}}}({\mathbf{y}})
\log_2\left(\frac{\hat{\mathcal{F}}({\mathbf{y}})}{m({\mathbf{y}})}\right)\,
, \label{Jayn2}\end{eqnarray}
\noi which precisely coincides with Jaynes's
finding~\cite{Jay1,Jay2}. Entropy (\ref{Jayn2}) is also known as
the Kullback--Leibler relative entropy.

\section*{Appendix F}

In this appendix we derive relation (\ref{mes6}).  To start we
must first identify ${\mathcal{E}}_n$. If we denote $N_l(a_i)$ as
the number of boxes of size $l$ needed to cover the unifractal
with the singularity exponent $a_i$  then ${\mathcal{E}}_n = \{
{\mathcal{E}}_{nk}(a_i); \ k \in N_l(a_i), i \in \mathbb{N} \}$.
Because of the scaling property we must set
${\mathcal{E}}_{nk}(a_i) = c_k(a_i) l^{a_i}$ with $c_k(a_i)$
weakly $l$ dependent. In order to
${\mathcal{I}}_{\alpha}({\mathcal{E}}_n)$ represent the ``ground
state" information we must require $c_k(a_i)$ to be a constant
(i.e., $c_k(a_i,l) = c(l)$). This is so because in such a case our
lack of information about the multifractal system (provided we
comply with the scaling of probability) is clearly highest. This
implies that $c = 1/\sum_i N_l(a_i) l^{a_i}$ as indeed
\begin{equation}
\sum_l {\mathcal{E}}_{n l} = \sum_i \sum_{k=1}^{N_l(a_i)}
{\mathcal{E}}_{nk}(a_i)=\sum_i N_l(a_i) \ c \ l^{a_i} = 1\, .
\end{equation}
\noi Notice that $c$ is weakly $l$ dependent since $ \sum_i
N_l(a_i) l^{a_i} \sim l^{\tau(1)} =1$. To proceed further we
employ the multifractal measure (\ref{mes2}). There
${\mathcal{P}}_n = \{ p_{nk} \}$ is the discrete (integrated)
probability distribution on the covering mesh. In case that the
limit in (\ref{mes2}) exists we may define the increment of
$\mu_{{\mathcal{P}}}^{(\alpha)}(d;l)$ between $a$ and $a+ da$ in
the small $l$ limit as
\begin{equation}
d\mu^{(\alpha)}_{{\mathcal{P}}}(a) = \lim_{l \rightarrow 0} \
\sum_{l^{a+ da} \stackrel{\mbox{~}_{\large{<}}}{\mbox{~}_{\sim}}\
p_{nk} \stackrel{\mbox{~}_{\large{<}}}{\mbox{~}_{\sim}} \ l^a }
\frac{p_{nk}^{\alpha}}{l^{\tau}}\, . \label{mes3}
\end{equation}
\noi Eq.(\ref{mes3}) then implies that
\begin{eqnarray}
\lim_{l\rightarrow 0} \ \log_2 \left(\sum_i \sum_{k}^{N_l(a_i)}
p_{nk}^{\alpha}(a_i)\right) &\approx& \log_2 \int_a
d\mu^{(\alpha)}_{{\mathcal{P}}}(a)\nonumber \\
&+& \tau(\alpha) \log_2 l\, ,
\end{eqnarray}
\noi and so especially
\begin{eqnarray}
\tilde{{\mathcal{I}}}_{\alpha}(\mu_{{\mathcal{P}}}) &\equiv&
\lim_{l \rightarrow 0} \
\left({\mathcal{I}}_{\alpha}({\mathcal{P}}_n) -
{\mathcal{I}}_{\alpha}({\mathcal{E}}_n) \right)\nonumber \\
&=& \frac{1}{(1-\alpha)} \ \log_2\left( \frac{\int_a
d\mu_{{\mathcal{P}}}^{(\alpha)}(a)}{\int_a
d\mu_{{\mathcal{E}}}^{(\alpha)}(a)}\right)\, . \label{mes4}
\end{eqnarray}
\noi Under the condition that the integrals exist relation
(\ref{mes4}) represents a well defined (and  finite) information
measure. From the same reasons as in Section III we may conclude
that $\tilde{\mathcal{I}}_{\alpha}(\mu_{\mathcal{P}})$ represents
negentropy. Notice that similarly as before
\begin{equation}
\left.\int_a d\mu_{{\mathcal{E}}}^{(\alpha)}(a)\right|_{V=1} =1\,
. \label{mes45}
\end{equation}
\noi This results from the fact that $\sum_i N_l(a_i) l^{\alpha
a_i - \tau(\alpha)}$ is $\alpha$ independent in the small $l$
limit. Actually,
\begin{eqnarray}
\frac{d}{d \alpha}K(\alpha) &\equiv&  \frac{d}{d \alpha} \sum_i
N_l (a_i) l^{\alpha a_i - \tau(\alpha)}
\nonumber \\
&=& \frac{d}{d\alpha}  \int da \ n(a) l^{-f(a) + \alpha a -
\tau(\alpha)}\nonumber \\
&=& \ln l  \int da (a - a_0) \ n(a) l^{-f(a)
+ \alpha a - \tau(\alpha)} \nonumber \\
&=& 0 + {\mathcal{O}}\left(     \frac{1}{(\ln l)^{3/2}}\right)\, ,
\label{mes5}
\end{eqnarray}
\noi On the last line of (\ref{mes5}) we have applied Laplace's
formula of the asymptotic calculus\cite{MoFe1}. Eq.(\ref{mes5})
confirms our previous assertion as it assures that the vanishing
of $dK(\alpha)/d\alpha$ at $l \rightarrow 0$ is at least as large
as that of $1/(\ln l)^{3/2}$. The consequence of this is that
\begin{eqnarray}
&&\lim_{l  \rightarrow 0} \ \frac{\sum_i N_l(a_i) l^{\alpha a_i -
\tau(\alpha) }}{\left( \sum_i N_l(a_i) l^{a_i} \right)^{\alpha}} =
\frac{K(\alpha)}{(K(1))^{\alpha}}  =
(K(1))^{1-\alpha}\nonumber \\
&& \mbox{\hspace{1cm}} = \frac{K(0)}{(K(1))^{\alpha}} =
\frac{V}{(K(1))^{\alpha}} = \frac{1}{(K(1))^{\alpha}}\, .
\end{eqnarray}
\noi The latter implies that $K(1) = 1$ and ergo (\ref{mes45})
holds.

\section*{Appendix G}

We show here an alternative way to obtain the real inverse formula
for Eq.(\ref{Lap1}). Let us start with the following observation:
\begin{eqnarray}
{\cal{F}}_{\cal{P}}(x) &=& \sum_{-\log_2 p_k < x} p_k = \sum_l p_l
\theta(\log_2 p_l + x)\, . \label{apC1}
\end{eqnarray}
\noi Using the limit representation of the step function
$\theta(x)$;
\begin{displaymath}
\theta(x) = \lim_{\varepsilon \rightarrow 0_+}
\exp\left(-2^{-\frac{x}{\varepsilon}} \right)\, ,
\end{displaymath}
\noi together with the functional relation
\begin{eqnarray}
\theta(\log_2 p_l + x) &=& \theta(x)\theta\left( \frac{\log_2
p_l}{x} + 1 \right)\nonumber \\
&+&  \theta(-x) \theta\left(- \frac{\log_2
p_l}{x} -1 \right) \nonumber \\
&=& \theta(x) - \varepsilon(x) \theta\left(- \frac{\log_2 p_l}{x}
-1 \right)\, ,
\end{eqnarray}
\noi we may rewrite (\ref{apC1}) as
\begin{equation}
{\cal{F}}_{\cal{P}}(x) = \theta(x) - \lim_{\varepsilon \rightarrow
0_+} \varepsilon(x) \sum_{n=0}^{\infty} \frac{(-1)^n
2^{\frac{n}{\varepsilon}}}{n!}\ 2^{-\frac{n}{\varepsilon x}
{\mathcal{I}}_{(n/\varepsilon x + 1)} }\, .
\end{equation}
\noi or equivalently
\begin{equation}
{\cal{F}}_{\cal{P}}^c(x) \approx \varepsilon(x) \sum_{n
=0}^{\infty} \frac{(-1)^n 2^{\Lambda n}}{n!}
2^{-\frac{n\Lambda}{x} {\mathcal{I}}_{(\Lambda n/x  + 1)}  }\, .
\end{equation}
\noi Here  the complementary information--distribution function of
${\cal{P}}$
\begin{displaymath}
{\cal{F}}_{\cal{P}}^{c}(x) \equiv \theta(x) -
{\cal{F}}_{\cal{P}}(x) = \sum_{-\log_{2} p_k \geq x \geq 0} p_k\,
,
\end{displaymath}
\noi was defined. The regulator $\Lambda \sim 1/\varepsilon$. Note
that because $x \in [0, + \infty)$ we have that $\alpha \in [1, +
\infty ) $. This is in the agreement with the analysis based on
the Widder--Stiltjes inverse formula.

\section*{Appendix H}

In this appendix we derive the reconstruction theorem for THC
entropy. Starting with Eq.(\ref{LST1}) we may write
\begin{eqnarray}
{\cal{F}}_{\cal{P}}(x) &=& \frac{1}{2\pi i} \int_{-i \infty +
\sigma}^{i \infty + \sigma} dp \;
\frac{e^{px} \ e^{-p{\cal{I}}_{\alpha}({\cal{P}})}}{p}\nonumber \\
&=& -\frac{1}{\ln(4) \pi i}\int_{-i \infty + \sigma}^{i \infty + \sigma}
dp \; e^{px} \, {\mathcal{S}}_{\alpha}({\mathcal{P}}) + \theta(x)
\,
\end{eqnarray}
\noi where the step function $\theta(x)$ was added and subtracted and
the Bromwich representation
\begin{displaymath}
\theta(x) = \frac{1}{2\pi i} \int_{-i \infty +\sigma}^{i \infty +
\sigma} dp \; \frac{e^{px}}{p}\, ,
\end{displaymath}
\noi was used. As a result we obtain
\begin{equation}
{\cal{F}}_{\cal{P}}^{c}(x) =
\frac{1}{\ln(4) \pi i}\int_{-i \infty + \sigma}^{i \infty + \sigma}
dp \; e^{px} \, {\mathcal{S}}_{\alpha}({\mathcal{P}})\, .
\end{equation}
\noi The inverse Laplace--Stiltjes transformation then gives
\begin{equation}
{\mathcal{S}}_{\alpha}({\mathcal{P}}) = \frac{1}{(\alpha
-1)}\int_{x=0}^{\infty}
 2^{(1-\alpha)x} \, d{\cal{F}}_{\cal{P
}}^{c}(x)\, .
\end{equation}


\end{document}